\documentclass[doublespacing]{elsart}
\usepackage{epsfig}

\usepackage{amssymb}
\usepackage{amsmath}

\begin{document}

\begin{frontmatter}



\title{Electron self-trapping at quantum and
classical critical points}


\author[1]{M. I. Auslender} and
\author[2]{M. I. Katsnelson}

\address[1]{Ben-Gurion University of the Negev, POB 653, Beer Sheva 84105, Israel}
\address[2]{Institute for Molecules and Materials, Radboud
University Nijmegen, 6525 ED, Nijmegen, The Netherlands}

\begin{abstract}
Using Feynman path integral technique estimations of the ground
state energy have been found for a conduction electron interacting
with order parameter fluctuations near quantum critical points. In
some cases only \textit{singular} perturbation theory in the
coupling constant emerges for the electron ground state energy. It
is shown that an autolocalized state (quantum fluctuon) can be
formed and its characteristics have been calculated depending on
critical exponents for both weak and strong coupling regimes. The
concept of fluctuon is considered also for the classical critical
point (at finite temperatures) and the difference between quantum
and classical cases has been investigated. It is shown that,
whereas the quantum fluctuon energy is connected with a true
boundary of the energy spectrum, for classical fluctuon it is just
a saddle-point solution for the chemical potential in the
exponential density of states fluctuation tail.
\end{abstract}

\end{frontmatter}

\section{Introduction}

The physics of quantum critical point (QCP) \cite
{hertz,sachdev,girvin,laughlin,vojta} is now a subject of growing
interest. There is a solid experimental evidence of relevance of
the QCP and related phenomena for ferroelectrics \cite{SrTiO},
high-temperature superconductors \cite{highTc1,highTc2},
Bose-Einstein condensed atoms in traps \cite{BEC}, itinerant
electron magnets \cite{SrRuO,YbRuSi,betaMn,MnSi}, heavy fermion
compounds \cite{coleman1,coleman2} and many other systems. Similar
to classical critical points or second-order phase transitions, a
scaling concept is of crucial importance near the QCP and
universal critical exponents can be introduced, which determine
all anomalous properties of the systems near QCP \cite{sachdev}.
The universality means that the basic physics depends not on the
details of a microscopic Hamiltonian but rather on space
dimensionality, dispersion law of low-frequency and
long-wavelength fluctuations of an order parameter and symmetry
properties of their effective action. In contrast with classical
phase transitions at finite temperatures thermodynamics of the QCP
is essentially dependent on the \textit{dynamical} critical
exponents \cite{hertz}.

There is an interesting issue how these critical fluctuations can
effect on the state of an excess charge carrier which appears as a
result of doping, injection, photoexcitation, etc. One can
consider for example the electron motion in a crystal near the
ferroelectric quantum phase transition in virtual ferroelectrics
such as SrTiO$_{3}$ or KTaO$_{3}$ under doping or pressure
\cite{SrTiO,blinc}, or near quantum magnetic phase transition due
to competing exchange interactions \cite{sachdev}. To our
knowledge this problem has not been considered yet. One may
speculate that a specific nature of the order parameter is not
very essential for this problem; due to softness and long-range
character of the critical fluctuations the effects of their
interaction with the conduction electrons may be very strong. In
particular, we will see that a self-trapping (autolocalization) of
the carrier proves possible, similar to a polaron formation in
ionic crystals \cite{VK,mahan} or spin polarons (``ferrons'') in
magnetic semiconductors \cite{nagaev}. A general concept of the
self-trapped electronic state due to interaction with order
parameter fluctuations (``fluctuon'') has been proposed many years
ago by Krivoglaz \cite{krivoglaz}. It appeared, however, that his
phenomenological approach is not applicable near the critical
point where the fluctuon radius is smaller than the correlation
length \cite{ourTMF}. We have considered this case \cite
{ourTMF,ourJMMM,ourDAN} using Feynman path integral variational
approach developed him for the polaron problem
\cite{feynman,path1}. Here we apply similar technique to consider
the quantum case. It will be shown that the classical and quantum
fluctuons are drastically different: if the latter can be
considered as a specific quasiparticle the former one represents
some quasilocalized state in the density of states tail. Apart
from possible applications to condensed matter physics the problem
under consideration gives a nontrivial example of the interaction
of a fermion with a bosonic quantum field with anomalous scaling
properties.

Whereas only the case of dispersionless Einstein phonon has been
considered originally by Feynman, later this method has been used
also to describe the interaction of electron with acoustic phonons
\cite{acous1,acous2}. We consider here a general case of
fluctuations with arbitrary dynamics which can be, in particular,
of dissipative type. The answers will be written in terms of some
frequency momenta of the fluctuations. One can assume that the
type of the fluctuation dynamics, being relevant, e.g., for
transport phenomena is not essential for static characteristics
such as autolocalization radius and energy; anyway, the method
used by us gives a rigorous upper limit for the ground-state
energy. Another difference (which is more important) is that the
phonon field is Gaussian whereas the Gaussian approximation for
the fluctuations which we will use can be justified only for not
too large coupling constants. It leads to some restrictions which
will be derived separately for all cases under consideration.

The interaction of electrons with quantum critical fluctuations
are intensively studied, especially in connection with
high-temperature superconductivity and heavy-fermion systems (for
review, see Refs. \cite{highTc1,highTc2} and
\cite{coleman1,coleman2}, respectively). Usually it is assumed
that the coupling constant is small in comparison with the Fermi
energy. Here we consider the case of {\it single} carrier where
the character of electron states is essentially different; one can
say that this difference is similar to the difference between
localized and extended states for the disordered systems. As a
next step, it would be interesting to consider degenerate gas of
fluctuons where the Fermi energy is finite but small in comparison
with the autolocalization energy which might be a subject of
future investigations.

The paper is organized as follows. In Section 2 we overview
general formalism for solving the problem posed. In the present
paper we consider the case of not too large coupling constant,
where the problem can be considered in Gaussian approximation for
the interaction with the fluctuations; explicit criteria are
presented below. The quantum case (zero temperature) is considered
in Section 3. Using the scaling properties of the fluctuation
spectral density (Subsection 3.1) we construct regular
perturbative expansion of the energy in the coupling constant
(weak-coupling regime, Subsection 3.2) as well as singular
perturbative expansion in strong coupling regime (Subsection 3.3).
The very existence of the regular perturbative regime depends
crucially on the value of dynamical critical exponent $z$ and
anomalous dimension $d$. The problem of fluctuon at classical
critical point (finite temperature) is treated in Section 4. We
solve the problem by both Feynman variational method (Subsection
4.1) and using Green function technique with vertex corrections
via Ward identity (Subsection 4.2). The similarity of the results
as regards dependencies of the density of states on the energy and
 coupling constant justifies the variational approach.

\section{Formulation of the problem using Feynman path integral}

For simplicity, we will consider the case of a scalar
order-parameter acting only on the orbital motion of the electron
and not on its spin (for example it may be the QCP in
ferroelectrics). Then, in continuum approximation, the Hamiltonian
of the system consisting of the electron and the order-parameter
field can be written in a simple form

\begin{equation}
\mathcal{H}=\mathcal{H}_{f}\left( \varphi \right)
+\mathcal{H}_{e}\left( \mathbf{r,}\varphi \right)
,\;\mathcal{H}_{e}\left( \mathbf{r,}\varphi \right)
=-\frac{1}{2}\nabla _{\mathbf{r}}^{2}-g\varphi \left( \mathbf{r}
\right)   \label{ham}
\end{equation}
where we have chosen the units $\hbar =m=1$, $m$ is the electron
effective mass, $\mathbf{r}$ is the electron coordinate, $\varphi
\left( \mathbf{r} \right) $ is the quantum order-parameter field
with its own Hamiltonian $\mathcal{H}_{f}\left( \varphi \right) $
and $g$ is the coupling constant. The partition function of the
whole system may be transformed to
\begin{equation}
Z=\mbox{Tr}e^{-\beta \mathcal{H}_{f}\left( \varphi \right) -\beta \mathcal{H}%
_{e}\left( \mathbf{r,}\varphi \right) }=Z_{f}\left\langle \mbox{Tr}_{\mathbf{%
r}}T_{\tau }\exp \left[ -\int_{0}^{\beta }\mathcal{H}_{e}\left(
\mathbf{r,}\varphi \left( \mathbf{r,}\tau \right) \right) d\tau
\right] \right\rangle _{f}  \label{partfun}
\end{equation}
where $Z_{f}=\mbox{Tr}_{\varphi }e^{-\beta \mathcal{H}_{f}\left(
\varphi
\right) }$ is the partition function of the field, $\varphi \left( \mathbf{r,%
}\tau \right) =e^{\tau \mathcal{H}_{f}\left( \varphi \right)
}\varphi \left( \mathbf{r}\right) e^{-\tau \mathcal{H}_{f}\left(
\varphi \right) }$ and
\begin{equation}
\left\langle \mathcal{A}\left( \varphi \right) \right\rangle _{f}=\frac{1}{%
Z_{f}}\mbox{Tr}_{\varphi }e^{-\beta \mathcal{H}_{f}\left( \varphi \right) }%
\mathcal{A}\left( \varphi \right)   \label{meanf}
\end{equation}
is the average over the field states. Using Feynman path-integral
approach \cite{path1,path2,path3} and taking average over
$\varphi$ yields for the electron-only free energy
\begin{equation}
\mathcal{F}=-\frac{1}{\beta }\left( \ln Z-\ln Z_{f}\right) =-\frac{1}{\beta }%
\ln \int_{\mathbf{r}\left( 0\right) =\mathbf{r}\left( \beta
\right) }e^{-\mathcal{S}}\mathcal{D}\left[ \mathbf{r}\left( \tau
\right) \right] , \label{freen}
\end{equation}
where $\mathcal{S}_{0}+\mathcal{S}_{int}$ is the effective action,
\begin{eqnarray}
\mathcal{S}_{0}&=&\frac{1}{2}\int_{0}^{\beta }\left[
\overset{\bullet }{\mathbf{r}}\left( \tau \right) \right]
^{2}d\tau \nonumber \\
\mathcal{S}_{int}&=&-\sum_{m=2}^{\infty }
\frac{g^{m}}{m!}\int_{0}^{\beta }...\int_{0}^{\beta }
\mathcal{K}_{m}\left( \mathbf{r}\left( \tau _{1}\right) ,\tau
_{1};...; \mathbf{r}\left( \tau _{m}\right) ,\tau _{m}\right)
d\tau _{1}...d\tau _{m} \label{efact}
\end{eqnarray}
and $\mathcal{K}_{m}\left( \mathbf{r}_{1},\tau
_{1};...;\mathbf{r}_{m},\tau _{m}\right) $ is the m-th cumulant
correlators, defined recursively by
\begin{align}
\mathcal{K}_{1}\left( \mathbf{r}_{1},\tau _{1}\right) &
=\left\langle \varphi \left( \mathbf{r}_{1}\mathbf{,}\tau
_{1}\right) \right\rangle _{f},
\notag \\
\mathcal{K}_{2}\left( \mathbf{r}_{1},\tau _{1};\mathbf{r}_{2},\tau
_{2}\right) & =\left\langle T_{\tau }\left[ \varphi \left( \mathbf{r}_{1}%
\mathbf{,}\tau _{1}\right) \varphi \left(
\mathbf{r}_{2}\mathbf{,}\tau
_{2}\right) \right] \right\rangle _{f}-\mathcal{K}_{1}\left( \mathbf{r}%
_{1},\tau _{1}\right) \mathcal{K}_{1}\left( \mathbf{r}_{2},\tau
_{2}\right) ,...  \label{cumulant}
\end{align}
etc. Further we will consider only the cases where
$\mathcal{K}_{1}=0$.

To estimate $\mathcal{F}$ and electron energy $\mathcal{E\,}%
=\lim_{\beta \rightarrow \infty }\,\mathcal{F}$ we use the same
trial action as was proposed by Feynman for the polaron problem
\cite{feynman} $\mathcal{S}_{t}=\mathcal{S}_{0}+\mathcal{S}_{pot}$
where
\begin{equation}
\mathcal{S}_{pot}=\frac{C}{2} \int_{0}^{\beta }\int_{0}^{\beta
}\left[ \mathbf{r}\left( \tau \right) -\mathbf{r}\left( \sigma
\right) \right] ^{2}e^{-w\left| \tau -\sigma \right| }d\tau
d\sigma, \label{tract}
\end{equation}
$C$ and $w$ being trial parameters. Then the
Peierls-Feynman-Bogoliubov inequality reads
\begin{equation}
\mathcal{F}\leq \mathcal{F}_{t}+\frac{1}{\beta }\left\langle \mathcal{S}%
_{int}-\mathcal{S}_{pot}\right\rangle _{t}  \label{PBineq}
\end{equation}
where
\begin{equation}
\mathcal{F}_{t}=-\frac{1}{\beta }\ln \int_{\mathbf{r}\left( 0\right) =%
\mathbf{r}\left( \beta \right)
}e^{-\mathcal{S}_{t}}\mathcal{D}\left[
\mathbf{r}\left( \tau \right) \right] ,\;\left\langle \mathcal{A}%
\right\rangle _{t}=\int_{\mathbf{r}\left( 0\right)
=\mathbf{r}\left( \beta \right) }\mathcal{A}\left[
\mathbf{r}\left( \tau \right) \right] e^{\beta
\mathcal{F}_{t}-\mathcal{S}_{t}}\mathcal{D}\left[ \mathbf{r}\left(
\tau \right) \right],  \label{trfun}
\end{equation}
which is equivalent to
\begin{align}
\mathcal{F}& \leq \mathcal{F}_{t}-\frac{C}{2\beta }\int_{0}^{\beta
}\int_{0}^{\beta }\left\langle \left[ \mathbf{r}\left( \tau \right) -%
\mathbf{r}\left( \sigma \right) \right] ^{2}\right\rangle
_{t}e^{-w\left|
\tau -\sigma \right| }d\tau d\sigma   \notag \\
& -\sum_{m=2}^{\infty }\frac{g^{m}}{m!\beta }\int_{0}^{\beta
}...\int_{0}^{\beta }\left\langle \mathcal{K}_{m}\left( \mathbf{r}%
\left( \tau _{1}\right) ,\tau _{1};...;\mathbf{r}\left( \tau
_{m}\right) ,\tau _{m}\right) \right\rangle
_{t}\prod_{j=1}^{m}d\tau _{j} \label{interim1}
\end{align}

To proceed, we will pass to the Fourier transforms
\begin{align}
& \left\langle \mathcal{K}_{m}\left( \mathbf{r}\left( \tau
_{1}\right) ,\tau _{1};...;\mathbf{r}\left( \tau _{m}\right) ,\tau
_{m}\right) \right\rangle
_{t}  \notag \\
& =\int ...\int \beta ^{1-m}\sum_{\omega _{1}...\omega _{m-1}}\mathcal{K}%
_{m}\left( \mathbf{K}_{1},i\omega
_{1};...;\mathbf{K}_{m-1},i\omega _{m-1}\right) \exp \left[
i\sum_{j=1}^{m-1}\omega _{j}\left( \tau
_{j}-\tau _{m}\right) \right]   \notag \\
& \times \left\langle \exp \left[ i\sum_{j=1}^{m-1}\mathbf{K}%
_{j}\cdot \left[ \mathbf{r}\left( \tau _{j}\right)
-\mathbf{r}\left( \tau
_{m}\right) \right] \right] \right\rangle _{t}\prod_{j=1}^{m-1}\frac{%
\Omega _{D}d^{D}K_{j}}{\left( 2\pi \right) ^{D}},\quad
\label{interim2}
\end{align}
where $\mathbf{K}_{j}$ are the wave-vectors, $\Omega _{D}$ is the
unit lattice cell volume, and $\omega _{j}$ are the bosonic
Matsubara frequencies. For the Gaussian trial action
$\mathcal{S}_{t}$ one has
\begin{equation}
\left\langle \exp \left\{ i\sum_{j=1}^{m-1}\mathbf{K}_{j}\cdot
\left[ \mathbf{r}\left( \tau _{j}\right) -\mathbf{r}\left( \tau
_{m}\right) \right] \right\} \right\rangle_{t}=\exp \left[
-\frac{1}{2}\sum_{j,k=1}^{m-1}f\left( \tau_{j}-\tau_{m},\tau
_{k}-\tau _{m}\right) \mathbf{K}_{j}\cdot \mathbf{K}_{k}\right],
\label{interim3}
\end{equation}
where
\begin{align}
& f\left( \tau _{j}-\tau _{m},\tau _{k}-\tau _{m}\right)  =\frac{1}{D}%
\left\langle \left[ \mathbf{r}\left( \tau _{j}\right)
-\mathbf{r}\left( \tau
_{m}\right) \right] \cdot \left[ \mathbf{r}\left( \tau _{k}\right) -\mathbf{r%
}\left( \tau _{m}\right) \right] \right\rangle _{t} =  \notag \\
& \frac{1}{2D}\left\{ \left\langle \left[ \mathbf{r}\left( \tau
_{j}\right) -\mathbf{r}\left( \tau _{m}\right) \right]
^{2}\right\rangle _{t}+\left\langle \left[ \mathbf{r}\left( \tau
_{k}\right) -\mathbf{r}\left(
\tau _{m}\right) \right] ^{2}\right\rangle _{t}-\left\langle \left[ \mathbf{r%
}\left( \tau _{j}\right) -\mathbf{r}\left( \tau _{k}\right)
\right] ^{2}\right\rangle _{t}\right\}
 \label{interim4}
\end{align}
Substituting Eqs.(\ref{Feyn2}), (\ref{interim2})-(\ref{interim4})
into Eq.(\ref{interim1}) we find an exact upper-bound estimation
for the free energy as a series in the coupling constant
\begin{align}
& \mathcal{F} \leq \mathcal{F}_t - \frac{1}{\beta}\left\langle
\mathcal{S}_{pot} \right\rangle- \notag \\
& \sum_{m=2}^{\infty }\frac{g^{m}}{\beta^{m} m!}\int
\int_{0}^{\beta }....\int \int_{0}^{\beta }\sum_{\omega
_{1}...\omega _{m-1}} \mathcal{K}_{m}\left( \mathbf{K}_{1},i\omega
_{1};..;\mathbf{K}_{m-1},i\omega _{m-1}\right) \times   \notag \\
& \exp \left\{ i\sum_{j=1}^{m-1}\omega _{j}\left( \tau _{j}-\tau
_{m}\right) -\frac{1}{2}\sum_{j,k=1}^{m-1} f\left( \tau _{j}-\tau
_{m},\tau _{k}-\tau _{m}\right) \mathbf{K}_{j}\cdot
\mathbf{K}_{k}\right\} \times \notag \\
& \prod_{j=1}^{m-1} \frac{\Omega _{D}d^{D}K_{j}d\tau _{j}}{\left(
2\pi \right) ^{D}} d\tau _{m}. \label{finest}
\end{align}

In this paper we restrict ourselves to Gaussian approximation,
which will mean \textit{ad hoc} the neglect of the cumulant terms
with $m>2$ in the series of Eq.(\ref{finest}). Unless $\varphi
\left( \mathbf{r,}\tau \right) $ is a Gaussian field indeed, the
Gaussian approximation is believed valid in a range of small
enough $g$, necessarily satisfying the condition
\begin{equation}
\frac{\left| g\right| }{W}\ll 1,  \label{w-bcond}
\end{equation}
where $W$ is a measure of the electron band width. Explicit
criterion for applicability of the Gaussian approximation depends
crucially on the critical exponents and space dimensionality, see
Section 3.

\section{Quantum case}

It was demonstrated by Feynman \cite{feynman} that at $\beta
\rightarrow \infty $
\begin{equation}
\frac{1}{D}\left\langle \left[ \mathbf{r}\left( \tau \right) -\mathbf{r}%
\left( \sigma \right) \right] ^{2}\right\rangle _{t}=\frac{v^{2}-w^{2}}{v^{3}%
}\left( 1-e^{-v\left| \tau -\sigma \right| }\right) +\frac{w^{2}}{v^{2}}%
\left| \tau -\sigma \right| ,\;v^{2}=w^{2}+\frac{4C}{w}
\label{Feyn1}
\end{equation}
($D$ is the space dimension) and so, with the notation
$\lambda=v/w$, we obtain
\begin{equation}
\mathcal{F}_{t}-\frac{1}{\beta}\left\langle \mathcal{S}_{pot}
\right \rangle = \frac{Dv\left( 1-\lambda \right) ^{2}}{4}
\label{Feyn2}
\end{equation}
and
\begin{align}
f\left( \tau _{j}-\tau _{m},\tau _{k}-\tau _{m}\right) & =
\frac{1-\lambda ^{2}}{2v}\left( 1-e^{-v\left| \tau _{j}-\tau
_{m}\right| }-e^{-v\left| \tau _{k}-\tau _{m}\right| }+e^{-v\left|
\tau _{j}-\tau
_{k}\right| }\right)   \notag \\
& +\frac{\lambda ^{2}}{2}\left( \left| \tau _{j}-\tau _{m}\right|
+\left| \tau _{k}-\tau _{m}\right| -\left| \tau _{j}-\tau
_{k}\right| \right). \; \label{ff}
\end{align}
Using Eqs.(\ref{Feyn1}),(\ref{Feyn2}),(\ref{ff}) and the Debye
approximation for integration over $\mathbf{K}$, to obtain
\begin{align}
\mathcal{F}& \leq \frac{Dv\left( 1-\lambda \right) ^{2}}{4}%
-g^{2}A_{D}\int_{0}^{K_{\max }}\sum_{\omega }\mathcal{K}_{2}\left(
K,i\omega \right) \times  \notag \\
& \left[ \int_{0}^{\beta }\cos \omega \tau \left( 1-\frac{\tau }{%
\beta }\right) e^{-\frac{1}{2}\lambda ^{2}K^{2}\tau -\frac{1-\lambda ^{2}}{2v%
}\left( 1-e^{-v\tau }\right) K^{2}}d\tau \right] K^{D-1}dK,
\label{Gaussest}
\end{align}
where $A_{D}=\frac{\Omega _{D}}{2^{D-1}\pi ^{\frac{1}{2}D}\Gamma
\left( \frac{1}{2}D\right) }$, $\Gamma \left( x\right) $ being the
gamma function and $K_{\max }$ is the Debye wave-number cutoff
satisfying $A_{D}K_{\max }^{D}=D$. For completing the limit $\beta
\rightarrow \infty$ in Eq.(\ref{Gaussest}) we use the method of
residues to sum over the Bose frequencies and employ the spectral
representation
\begin{equation}
\mathcal{K}_{2}\left( K,i\omega \right) =\frac{1}{\pi
}\int_{-\infty }^{\infty }\frac{\mathcal{J}\left( K,u\right)
}{u-i\omega }du, \label{spectrepr}
\end{equation}
with $\mathcal{J}\left( K,x\right)$ being an appropriate spectral
density. So we obtain the variational upper-bound estimation
$\mathcal{E}\leq \mathcal{E}_{0}\left( v,\lambda \right) $, where
\begin{align}
& \mathcal{E}_{0}\left( v,\lambda \right) =\frac{Dv\left(
1-\lambda \right) ^{2}}{4}-
\notag \\
& \frac{g^{2}A_{D}}{2\pi }\int_{0}^{K_{\max }}\int_{0}^{\infty
}\int_{0}^{\infty }\mathcal{J}_{-}\left( K,u\right) e^{-\left( u+%
\frac{\lambda ^{2}}{2}K^{2}\right) t-\frac{1-\lambda
^{2}}{2v}\left( 1-e^{-vt}\right) K^{2}}K^{D-1}dKdudt
\label{fin-1}
\end{align}
and $\mathcal{J}_{-}\left( K,u\right) =\mathcal{J}\left(
K,u\right) - \mathcal{J}\left( K,-u\right)$. Note that for
even-frequency spectrum fluctuations (in particular static ones)
$\mathcal{J}_{-}\left( K,u\right) \equiv 0$, so the interaction
term in Eq.(\ref{fin-1}) vanishes.

\subsection{The use of scaling}

Until now the statistical properties of the field $\varphi \left( \mathbf{%
r,\tau }\right) $ have not been specified. Further we will use the
dynamical scaling law near the QCP \cite{sachdev}
\begin{equation}
\mathcal{J}_{-}\left( K,u\right) =f^{2-\eta }\mathcal{J}_{-}\left(
fK,f^{z}u\right) ,\;\forall f>0  \label{scaling}
\end{equation}
where $\eta $ and $z$ is an ``anomalous-dimension''\ and dynamical
critical exponent, respectively. Using in Eq.(\ref{scaling})
$f=K_{\max }/K$, we have
\begin{equation}
\mathcal{J}_{-}\left( K,u\right) =\left( \frac{K}{K_{\max
}}\right) ^{\eta -2}\mathcal{J}_{-}\left( K_{\max },\left(
\frac{K_{\max }}{K}\right) ^{z}u\right) .  \label{scaling1}
\end{equation}
Plugging Eq.(\ref{scaling1}) into Eq.(\ref{fin-1}) and using the
substitutions for the integration variables
\begin{equation}
u=\left( \frac{K}{K_{\max }}\right) ^{z}\varpi ,\,K=K_{\max }\sqrt{x}%
,\,t=v^{-1}s,  \label{subst-1}
\end{equation}
notations for the parameters
\begin{equation}
W=\frac{1}{2}K_{\max }^{2},\,q=\frac{v}{W},\,d=D-2+\eta ,
\label{not-1}
\end{equation}
$W$ being just the band width in the Debye approximation, and for
the function
\begin{equation}
\phi \left( s,\lambda ^{2}\right) =\left( 1-\lambda ^{2}\right)
\left( 1-e^{-s}\right) +\lambda ^{2}s,  \label{fun-phi1}
\end{equation}
as well as rescaling $g$ to fix the normalization of the
fluctuation spectrum
\begin{equation*}
\int_{0}^{\infty }Q\left( \varpi \right) d\varpi =1,\,Q\left(
\varpi \right) =\frac{1}{\pi }\widetilde{\mathcal{J}}\left(
K_{\max },\varpi \right) ,
\end{equation*}
we obtain
\begin{equation}
\mathcal{E}_{0}\left( v,\lambda \right) =\frac{D}{4}Wq\left(
1-\lambda \right)
^{2}-\frac{D}{4}\frac{g^{2}}{W}q^{-1}\left\langle
\int_{0}^{1}\int_{0}^{\infty }x^{\frac{d+z}{2}-1}e^{-q^{-1}%
\left[ \phi \left( s,\lambda ^{2}\right) x+\frac{\varpi }{W}sx^{\frac{z}{2}}%
\right] }dxds\right\rangle _{\varpi }  \label{fin-2}
\end{equation}
where the indexed by $\varpi $ angular brackets mean averaging
with the weight $Q\left( \varpi \right) $.

\subsection{Weak-coupling regime}

In the weak-coupling regime $\mu =1-\lambda \ll 1$, while the
range of the parameter $q$ is not predetermined yet (however, the
restriction $q<1$ should be imposed anyway, otherwise the
continuum description could not be used). In this regime the
electron is weakly ``fluctuation-dressed''. Using assumed
smallness of $\mu $ we can expand the right-hand side of
Eq.(\ref{fin-2}) in the Taylor series with respect to $\mu $. This
gives up to the terms of second order in $\mu$ inclusive
\begin{equation}
\mathcal{E}_{0}\left( v,\lambda \right) \simeq
-\frac{D}{4}a_{0}\left( d,z\right)
\frac{g^{2}}{W}-\frac{D}{2}a_{1}\left( d,z,q\right) \frac{g^{2}}
{W}\mu +\frac{D}{4}\left[ q+\left( \frac{g}{W}\right)
^{2}a_{2}\left( d,z,q\right) \right] W\mu ^{2}  \label{fin-3}
\end{equation}
where
\begin{equation}
a_{0}\left( d,z\right) =\left\langle \int_{0}^{1}\frac{x^{\frac{d+z}{2%
}-1}}{x+\frac{\varpi }{W}x^{\frac{z}{2}}}dx\right\rangle _{\varpi
}, \label{const-a_0}
\end{equation}
\begin{equation}
a_{1}\left( d,z,q\right) =q\left\langle \int_{0}^{1}\frac{x^{\frac{d+z%
}{2}}}{\left( x+\frac{\varpi }{W}x^{\frac{z}{2}}\right) ^{2}\left( q+x+\frac{%
\varpi }{W}x^{\frac{z}{2}}\right) }dx\right\rangle _{\varpi }
\label{fun-a_1}
\end{equation}
and
\begin{align}
a_{2}\left( d,z,q\right) & = 4q^{2}\left\langle
\int_{0}^{1}\frac{\left(
2q+3x+3\frac{\varpi }{W}x^{\frac{z}{2}}\right) x^{\frac{d+z}{2}+1}}{\left( x+%
\frac{\varpi }{W}x^{\frac{z}{2}}\right) ^{3}\left( q+x+\frac{\varpi }{W}x^{%
\frac{z}{2}}\right) ^{2}\left( 2q+x+\frac{\varpi }{W}x^{\frac{z}{2}}\right) }%
dx\right\rangle _{\varpi }
\notag \\
& -a_{1}\left(d,z,q\right)\label{fun-a_2}
\end{align}

The first term in Eq.(\ref{fin-3}) is the electron band edge shift
in the lowest-order Born approximation, the second term is the
potential energy and the third term is the renormalized kinetic
energy.

The Eq.(\ref{fin-3}) is to be minimized with respect to $\mu $ and
$q$. Let the optimum values of the variational parameters be $\mu
_{0}$ and $q_{0}$. Within the small $\mu $ regime, the correction
$\propto g^{2}$ to the bare kinetic energy that describes the
fluctuation-driven renormalization of the electron effective mass,
results in a contribution $\propto g^{6}$ to the optimal bound
$\mathcal{E}_{0}$. This contribution is negligible when expanding
$\mathcal{E}_{0}$ up to terms $\propto g^{4}$ inclusive. The
condition that allows to neglect the above renormalization reads
\begin{equation}
\left( \frac{g}{W}\right) ^{2}\frac{\left| a_{2}\left(
d,z,q_{0}\right) \right| }{q_{0}}\ll 1,  \label{cond1}
\end{equation}
which is, in general, consistent with Eq.(\ref{w-bcond}). Assuming
the condition of Eq.(\ref{cond1}) to hold, we minimize
Eq.(\ref{fin-3}) first in $\mu $ and next in $q$. This gives the
following expression for $\mu _{0}$ and $\mathcal{E}_{0}$
\begin{equation}
\mu _{0}=\left( \frac{g}{W}\right) ^{2}\frac{a_{1}\left( d,z,q_{0}\right) }{%
q_{0}}  \label{mu_0}
\end{equation}
and
\begin{equation}
\mathcal{E}_{0}=-\frac{D}{4}a_{0}\left( d,z\right) \frac{g^{2}}{W}-\frac{D}{4%
}a_{1}\left( d,z\right) \frac{g^{4}}{W^{3}},  \label{E_01}
\end{equation}
respectively, where the positive number $a_{1}\left( d,z\right) $
is the maximum of the function
\begin{equation}
G\left( d,z,q\right) =\frac{a_{1}^{2}\left( d,z,q\right) }{q},
\label{fun-G}
\end{equation}
viz.
\begin{equation}
a_{1}\left( d,z\right) =\max_{0<q<\infty }G\left( d,z,q\right)
=G\left( d,z,q_{0}\right) ,  \label{const-a_1}
\end{equation}
and $q_{0}$ is the point where this maximum is attained. Note that $%
\lim_{q\rightarrow \infty }G\left( d,z,q\right) =0$ due to
Eqs.(\ref {fun-a_1}), (\ref{fun-G}), so for existence of the above
maximum it would be sufficient that $\lim_{q\rightarrow 0}G\left(
d,z,q\right) =0$. As deduced from the very structure of
$\mathcal{E}_{0}\left( v,\lambda \right) $ (Eq.(\ref{fin-3})), the
parameter
\begin{equation}
l_{0}=\frac{1}{K_{\max }\mu _{0}\sqrt{q_{0}}}=\left( \frac{W}{g}\right) ^{2}%
\left[ a_{1}\left( d,z\right) \right] ^{-\frac{1}{2}}K_{\max
}^{-1} \label{fluct-size1}
\end{equation}
is a measure of the fluctuon potential-well size, which should be
much larger than the lattice constant, i.e. satisfy $l_{0}K_{\max
}\gg 1$.

By the virtue of Eq.(\ref{fun-a_2}) $\left| a_{2}\left(
d,z,q\right) \right|
>a_{1}\left( d,z,q\right) $. Therefore, once Eq.(\ref{cond1}) is checked to
hold, it automatically results in $\mu _{0}\ll 1$, due to
Eq.(\ref{mu_0}). On the other hand, inability to satisfy
Eq.(\ref{cond1}) would mean inapplicability of the perturbational
regime. After this general analysis, let us consider different
cases regarding the critical exponent $z$.

\subsubsection{The cases with $z\geq2$}

In this case we always have $\frac{\varpi }{W}x^{\frac{z}{2}}\ll
x$, due to smallness of non-adiabaticity parameter $\frac{\varpi
}{W}$, so Eqs.(\ref {const-a_0}) - (\ref{fun-a_2}) reduce to the
functions of the combined index $d^{\ast }=d+z-2$
\begin{equation}
a_{0}\left( d,z\right) \simeq A_{0}(d^{\ast}) =
\frac{2}{d^{\ast}}, \label{const-a_0zgeq2}
\end{equation}
\begin{equation}
a_{1}\left( d,z,q\right) \simeq A_{1}(d^{\ast},q) =
q^{\frac{d^{\ast }}{2}}\Phi _{d^{\ast }}\left( q\right),
\label{fun-a_1zgeq2}
\end{equation}
where
\begin{equation}
\Phi _{b}\left( x\right) =\int_{0}^{x^{-1}}\frac{t^{\frac{b}{2}-1}}{%
t+1}dt,\;b>0  \label{fun-Phi}
\end{equation}
and
\begin{align}
a_{2}\left( d,z,q\right) & \simeq A_{2}\left( d^{\ast },q\right) &
=\left( 11+2d^{\ast }\right) A_{1}\left( d^{\ast },q\right)
-8A_{1}\left( d^{\ast },2q\right) -\frac{4q}{1+q}.
\label{fun-a_2zgeq2}
\end{align}
The necessary condition for the finiteness of the above integrals
is $d^{\ast }>0$. One can see that in this case the fluctuation
spectral density shape is completely irrelevant.

To infer on existence of the maximum of $G\left( d,z,q\right)
=G_{d^{\ast
}}\left( q\right) =q^{d^{\ast }-1}\left[ \Phi _{d^{\ast }}\left( q\right) %
\right] ^{2}$ we first note that $\lim_{q\rightarrow 0}G_{d^{\ast
}}\left( q\right) =0$ at $2>d^{\ast }>1$, since for such $d^{\ast }$%
\begin{equation}
\lim_{q\rightarrow 0}\Phi _{d^{\ast }}\left( q\right)=\Gamma
\left( \frac{d^{\ast }}{2}\right) \Gamma \left( 1-\frac{d^{\ast
}}{2}\right) =\frac{\pi }{\sin \left( \frac{\pi d^{\ast
}}{2}\right) }.  \label{lim-Phi}
\end{equation}
For $d^{\ast }=2$, $\Phi _{2}\left( q\right) =\ln \left(
q^{-1}+1\right) $, and $\lim_{q\rightarrow 0}G_{2}\left( q\right)
=\lim_{q\rightarrow 0}q\ln ^{2}\left( q^{-1}+1\right) =0$ also.
The
functions $\Phi _{d^{\ast }}\left( q\right) $ for (rather unrealistic) case $%
4\geq d^{\ast }>2$ are reduced to those with $d^{\ast }\leq 2$
using the functional relation
\begin{equation}
q^{\frac{d^{\ast }}{2}}\Phi _{d^{\ast }}\left( q\right) =q\left[ \frac{2}{%
d^{\ast }-2}-q^{\frac{d^{\ast }-2}{2}}\Phi _{d^{\ast }-2}\left( q\right) %
\right] ,\;d^{\ast }>2  \label{funrel-Phi}
\end{equation}
and again we get $\lim_{q\rightarrow 0}G_{d^{\ast }}\left(
q\right) =0 $. As outlined in the previous subsection, this means
that at least one maximum point $0<q_{0}<\infty $ does exist at
$d^{\ast }>1$. One the other hand, the equation
$\frac{d}{dq}G_{d^{\ast }}\left( q\right) =0$ for determining
$q_{0}$ is rigorously transformed to the following one:
\begin{equation}
\left( d^{\ast }-1\right) \Phi _{d^{\ast }}\left( q\right) -\frac{2q^{\frac{%
2-d^{\ast }}{2}}}{q+1}=0,  \label{q_0-equat}
\end{equation}
which obviously has no solution if $d^{\ast }\leq 1$ . Thus for
$d^{\ast }\leq 1$, the weak coupling regime never applies. This
exponents range will be revisited  in Section 3.2.

For $1<d^{\ast }<2$, the assumption of small $q_{0}$ would allow
one, by the virture of Eq.(\ref{lim-Phi}), to solve approximately
Eq.(\ref{q_0-equat}) in a closed form. However, compared with
numerics for specific $d^{\ast }$, this approximation seems to be
too inaccurate. An approximate equation, which results from
inclusion of the next-to-leading terms of that asymptotic, can not
be solved analytically anymore. So given $d^{\ast }$, a reliable
calculation of $q_{0}$ requires numerical approach. For some cases
of rational $d^{\ast }$, one of them is considered below, $\Phi
_{d^{\ast }}\left( q\right)$ is expressed in elementary functions.

Let us put $d^{\ast}=\frac{3}{2}$. This case is a representative
for fractional-rational $d^{\ast }$. We have
\begin{align*}
A_{1}\left( \frac{3}{2},q\right) & =\sqrt{2}q^{\frac{3}{4}}\left[ \ln \frac{%
\left( 1+q^{-1}\right) ^{\frac{1}{2}}}{q^{-\frac{1}{2}}+\sqrt{2}q^{-\frac{1}{%
4}}+1}+\arctan \left( \sqrt{2}q^{-\frac{1}{4}}+1\right)
+\allowbreak \arctan
\left( \sqrt{2}q^{-\frac{1}{4}}-1\right) \right] , \\
G_{\frac{3}{2}}\left( q\right) & =2q^{\frac{1}{2}}\left[ \ln
\frac{\left(
1+q^{-1}\right) ^{\frac{1}{2}}}{q^{-\frac{1}{2}}+\sqrt{2}q^{-\frac{1}{4}}+1}%
+\arctan \left( \sqrt{2}q^{-\frac{1}{4}}+1\right) +\allowbreak
\arctan \left( \sqrt{2}q^{-\frac{1}{4}}-1\right) \right] ^{2}.
\end{align*}
The graph of $G_{\frac{3}{2}}\left( q\right) $ is shown in Fig.1.
Eq.(\ref {q_0-equat}) for $d^{\ast }=\frac{3}{2}$ has unique
solution $q_{0}\simeq 0.12\allowbreak 6$, for which
$G_{\frac{3}{2}}\left( q_{0}\right) =a_{1}\left(
\frac{3}{2}\right) \simeq 1.589.$ Checking Eq.(\ref{cond1}) yields
after cumbersome calculations
\begin{equation}
\frac{\left| g\right| }{W}\ll 0.378\,,  \label{cond1d*eq1.5}
\end{equation}
Provided that Eq.(\ref{cond1d*eq1.5}) holds, we obtain from
Eq.(\ref{E_01})
\begin{equation}
\mathcal{E}_{0}\simeq -\frac{D}{3}\frac{g^{2}}{W}\left( 1+1.\,\allowbreak 19%
\frac{g^{2}}{W^{2}}\right) ,  \label{E_01d*eq1.5}
\end{equation}
and from Eq.(\ref{fluct-size1})
\begin{equation}
l_{0}K_{\max }\simeq 0.793\left( \frac{W}{g}\right) ^{2}.
\label{l_0d*eq1.5}
\end{equation}

\begin{figure}
\includegraphics[width=\columnwidth]{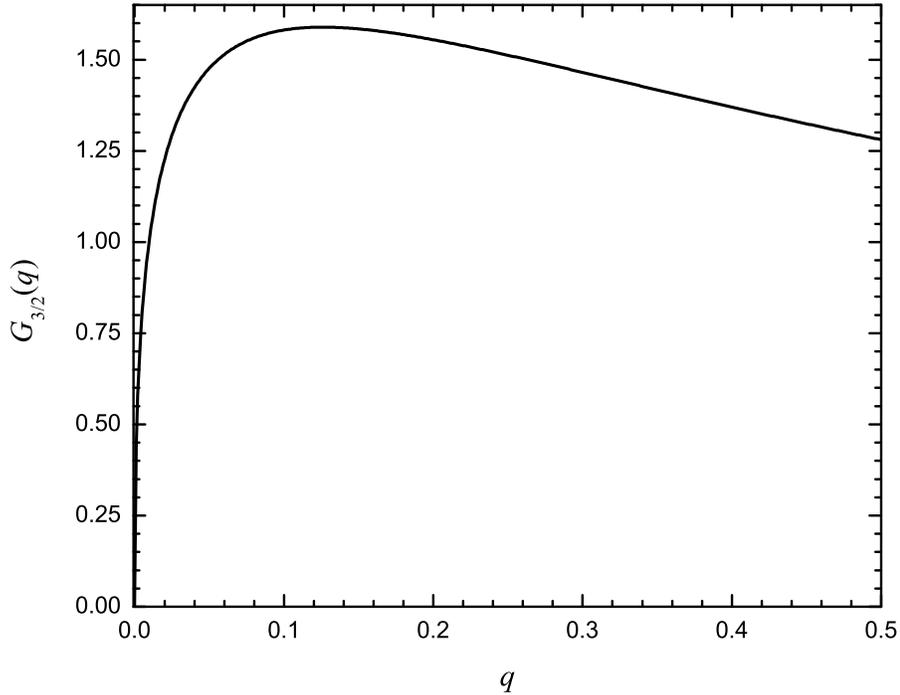}
\caption{\label{fig:epsart} Graph of the function
$G_{\frac{3}{2}}\left( q\right)$}
\end{figure}

The numerical results obtained for different values of
$d^{\ast}>1$ show that within the weak copling regime the smaller
$d^{\ast}$ the larger numerical factor of the fourth-order
correction in $\mathcal{E}_{0}$, and the narrower the range of $g$
where that approximation works.

\subsubsection{The cases with $0\leq z<2$}

For $0\leq z<2$, Eqs.(\ref{const-a_0}) - (\ref{fun-a_2}) are
transformed quite specifically. Let $\varpi _{0}$ scales
fluctuation frequencies, so that
$Q\left( \varpi \right) $ be a function of the reduced frequency $\nu =\frac{%
\varpi }{\varpi _{0}}$. Then, omitting from now on the index of
the averaging over $\varpi $ (or $\nu $), we have
\begin{equation}
a_{0}\left( d,z\right) =\left( \frac{W}{\varpi _{0}}\right)
^{1-\frac{d}{2-z}}\frac{2}{2-z} \left\langle \nu
^{\frac{d+z-2}{2-z}}\Phi _{\frac{2d}{2-z}}\left( \nu \frac{ \varpi
_{0}}{W}\right) \right\rangle  \label{const-a_0zl2}
\end{equation}
and $\Phi _{b}\left( x\right)$ is defined by Eq.(\ref{fun-Phi}).
It is seen that in the present case the weak coupling regime has a
sense only at $d>0$.

The asymptotic of $a_{0}\left( d,z\right) $ at $\frac{\varpi
_{0}}{W}\ll 1$ depends critically upon the sign of $d+z-2$,
yielding
\begin{equation}
a_{0}\left( d,z\right) \simeq \left\{
\begin{array}{r}
\frac{2}{2-z}\frac{\pi \left\langle \nu ^{\frac{d}{2-z}-1}\right\rangle }{%
\sin \left( \pi \frac{d}{2-z}\right) }\left( \frac{W}{\varpi
_{0}}\right)
^{1-\frac{d}{2-z}},\; \; d+z-2<0 \\
\frac{2}{d}\ln \left( \frac{W}{\varpi _{0}\overline{\nu }}\right)
,\; \; d+z-2=0
\\
\frac{2}{d+z-2},\; \; d+z-2>0
\end{array}
\right. ,  \label{con-a_0-zl2asym}
\end{equation}
where $\ln \overline{\nu }=\left\langle \ln \nu \right\rangle $
and Eq.(\ref {lim-Phi}) is taken into account. Thus, the Born
energy scale depends on the fluctuation dynamics: (i) drastically
in the first subcase, including in particular original Feynman's
polaron \cite{feynman}; (ii) weakly in the second subcase; (iii)
negligibly in the last subcase.

Next two integrals (\ref{fun-a_1}),(\ref{fun-a_2}) are transformed
and asymptotically represented at $\frac{W}{\varpi _{0}}\gg 1$ as
follows
\begin{equation}
a_{i}\left( d,z,q\right) \simeq \left( \frac{W}{\varpi _{0}}\right) ^{1-%
\frac{d}{2-z}}A_{i}\left( d,z,\varkappa \right) ,\;i=1,2,
\label{fun-a_12-zl2asym}
\end{equation}
where $\varkappa =q\left( \frac{W}{\varpi _{0}}\right)
^{\frac{2}{2-z}}$ is a new variable to optimize over, and
\begin{equation}
A_{1}\left( d,z,\varkappa \right) =\frac{2\varkappa
}{2-z}\left\langle \int_{0}^{\infty
}\frac{u^{\frac{d}{2-z}}du}{\left( u+\nu \right)
^{2}\left( \varkappa +u^{\frac{2}{2-z}}+\nu u^{\frac{z}{2-z}}\right) }%
\right\rangle ,  \label{fun-A_1}
\end{equation}
\begin{align}
A_{2}\left( d,z,\varkappa \right) & =\frac{8\varkappa
^{2}}{2-z}\left\langle
\int_{0}^{\infty }\frac{\left( 2\varkappa +3u^{\frac{2}{2-z}}+3\nu u^{%
\frac{z}{2-z}}\right) u^{\frac{d}{2-z}+1}du}{\left( u+\nu \right)
^{3}\left( \varkappa +u^{\frac{2}{2-z}}+\nu
u^{\frac{z}{2-z}}\right) ^{2}\left( 2\varkappa
+u^{\frac{2}{2-z}}+\nu u^{\frac{z}{2-z}}\right) }\right\rangle
\notag \\
& -A_{1}\left( d,z,\varkappa \right) .  \label{fun-A_2}
\end{align}
Note that for any reasonable $d$ the integrands in Eqs.(\ref{fun-A_1}) and (%
\ref{fun-A_2}) fall off at $u\rightarrow \infty $ faster than
$u^{-2}$. Therefore, in the both integrals, unlike that in
Eq.(\ref{const-a_0zl2}),
the upper limit $\frac{W}{\varpi _{0}}\gg 1$ has been safely replaced by $%
\infty $.

In the case considered the expression (\ref{fun-G}) is
parametrized as follows
\begin{equation}
G\left( d,z,q\right) =\left( \frac{W}{\varpi _{0}}\right) ^{2\frac{3-z-d}{2-z%
}}\overline{G}\left( d,z,\varkappa \right) ,\;\overline{G}\left(
d,z,\varkappa \right) =\frac{A_{1}^{2}\left( d,z,\varkappa \right) }{%
\varkappa }.  \label{fun-Gb}
\end{equation}
Accordingly, the energy asymptotic in weak coupling regime at
$z<2$ is given by
\begin{equation}
\mathcal{E}_{0}=-\frac{D}{4}\frac{g^{2}}{W}\left( \frac{W}{\varpi _{0}}%
\right) ^{\frac{2-z-d}{2-z}}\left[ \frac{2}{2-z}\frac{\pi \left\langle \nu ^{%
\frac{d}{2-z}-1}\right\rangle }{\sin \left( \frac{\pi d}{2-z}\right) }%
+\left( \frac{g}{W}\right) ^{2}\left( \frac{W}{\varpi _{0}}\right) ^{\frac{%
4-z-d}{2-z}}A_{1}\left( d,z\right) \right]  \label{E_0-dl2-z}
\end{equation}
for $z<2-d$,
\begin{equation}
\mathcal{E}_{0}=-\frac{D}{4}\frac{g^{2}}{W}\left[ \frac{2}{d}\ln
\left( \frac{W}{\varpi _{0}\overline{\nu }}\right) +\left(
\frac{g}{W}\right)
^{2}\left( \frac{W}{\varpi _{0}}\right) ^{\frac{2}{d}}A_{1}\left( d,z\right) %
\right]  \label{E_0-deq2-z}
\end{equation}
for $z=2-d$, and
\begin{equation}
\mathcal{E}_{0}=-\frac{D}{4}\frac{g^{2}}{W}\left[
\frac{2}{d+z-2}+\left(
\frac{g}{W}\right) ^{2}\left( \frac{W}{\varpi _{0}}\right) ^{2\frac{3-z-d}{%
2-z}}A_{1}\left( d,z\right) \right] ,  \label{E_0-dg2-z}
\end{equation}
for $z>2-d$, where
\begin{equation}
A_{1}\left( d,z\right) =\max_{0<\varkappa <\infty
}\overline{G}\left( d,z,\varkappa \right)  \label{const-A_1}
\end{equation}
Finally, Eq.(\ref{fluct-size1}) yields for the fluctuon size in
the present case
\begin{equation}
l_{0}K_{\max }\simeq \left( \frac{W}{g}\right) ^{2}\left( \frac{\varpi _{0}}{%
W}\right) ^{\frac{3-z-d}{2-z}}\left[ A_{1}\left( d,z\right) \right] ^{-\frac{%
1}{2}}.  \label{fluct-size-dl2-z}
\end{equation}
For a given $g$, the fluctuon size at $z<2$ proves parametrically
much smaller than that at $z\geq 2$ unless $d+z\geq 3$.

The sufficient condition for the perturbational regime to hold is
provided by Eq.(\ref{cond1}), which reads in the present case
\begin{equation}
\frac{\left| A_{2}\left( d,z,\varkappa _{0}\right) \right| }{\varkappa _{0}}%
\left( \frac{g}{W}\right) ^{2}\left( \frac{W}{\varpi _{0}}\right) ^{\frac{%
4-z-d}{2-z}}\ll 1,  \label{cond1zl2}
\end{equation}
where the numerical factor requires a numerical calculation for
specific $d$ and $z$. This condition proves much more stringent
than Eq.(\ref{w-bcond}), but assures that $l_{0}K_{\max }\gg 1$ in
any case.

Now the key question is that of existing the optimal $\varkappa
_{0}$, to answer which exploring the behavior of $A_{1}\left(
d,z,\varkappa \right) $ at $\varkappa \rightarrow 0$ is crucial.
Let us assume that $\left\langle
\nu ^{-\alpha }\right\rangle <\infty $ for all $\alpha >0$. Then at $z<1+%
\frac{1}{2}d$
\begin{equation*}
\lim_{\varkappa \rightarrow 0}\frac{A_{1}\left( d,z,\varkappa \right) }{%
\varkappa } =\frac{2-d}{\left( 2-z\right) ^{2}}\left\langle \nu ^{-\frac{%
4-d-z}{2-z}}\right\rangle \frac{\left( \pi \frac{d-z}{2-z}\right)
}{\sin \left( \pi \frac{d-z}{2-z}\right) }<\infty ,
\end{equation*}
so $\lim_{\varkappa \rightarrow 0}\overline{G}\left( d,z,\varkappa
\right) =0$. At $z=$ $1+\frac{1}{2}d$ we obtain the asymptotic at
$\varkappa \rightarrow 0$
\begin{align*}
A_{1}\left( d,z,\varkappa \right) & \approx
\frac{4}{2+d}\left\langle \nu ^{-3}\right\rangle \varkappa \ln \frac{%
\overline{\nu }^{\frac{4}{2+d}}\exp \left( -\frac{3}{2}\frac{2-d}{2+d}%
\right) }{\varkappa },\;\ln \overline{\nu }=\frac{\left\langle \nu
^{-3}\ln \nu \right\rangle }{\left\langle \nu ^{-3}\right\rangle
},
\end{align*}
so we have $\lim_{\varkappa \rightarrow 0}\overline{G}\left(
d,z,\varkappa \right) =0$ also in this case. Hence at $z\leq
1+\frac{1}{2}d$ the maximum point $\varkappa _{0}$ surely exists.
At $z>$ $1+\frac{1}{2}d$,
making use of the replacement $u=\left( \varkappa \nu ^{-1}\right) ^{\frac{%
2-z}{z}}t^{\frac{2-z}{z}}$ and of Eq.(\ref{lim-Phi}), we obtain
the following asymptotic
\begin{align*}
A_{1}\left( d,z,\varkappa \right)
& \simeq \frac{2}{z}\varkappa ^{\frac{d+2-z}{z}}\left\langle \nu ^{-\frac{%
d+2+z}{z}}\right\rangle \frac{\pi }{\sin \left( \pi \frac{d+2-z}{z}\right) }%
,\;\varkappa \rightarrow 0,
\end{align*}
from which we infer that $\varkappa _{0}$ exists, since $\lim_{%
\varkappa \rightarrow 0}\overline{G}\left( d,z,\varkappa \right) =0$, if $z<%
\frac{2}{3}\left( d+2\right) $ (that holds authomatically for
$d\geq 1$). If $z\geq \frac{2}{3}\left( d+2\right) $, which may
occur for $0<d<1$, the above limit is either a finite number or
$\infty $ that makes weak coupling regime nonexistent.

For completeness, it is instructive to consider numerical
examples. We consider two important cases $z=0$ and $z=1$ falling
into the class $z<1+ \frac{1}{2}d$, for which the existence of
$\varkappa_{0}$ has been proved above. In the both cases the
relevant formulas, before the $\nu$ averaging, are expessed in
elementary functions. Due to persisting $\nu$ averaging and
arbitrary $d$, however, the formulas yet remain too comlex for
illustrative numerics. To make things simpler, in the subsequent
two examples we assume that $d=1$ and the $\nu$ distribution is
strongly peaked at $\nu=1$. We do not expose the corresponding
graphs of $\overline{G}\left( 1,z,\varkappa\right) $ since they
are pretty much similar in shape to the graph shown in Fig.1,
apart of appreciable difference in scales of variables $\varkappa$
and $q$.

\paragraph*{Example: $d=1$, $z=0$.}

With the above assumption this is actually the Feynman polaron
problem \cite{feynman,path1}. We obtain
\begin{equation}
\overline{G}\left( 1,0,\varkappa \right) =\frac{\pi
^{2}}{\varkappa }\left( \frac{1}{2}+\frac{1-\sqrt{1+\varkappa
}}{\varkappa }\right) ^{2}. \label{fun-Gbdeq1zeq0}
\end{equation}
This function achieves its maximum at $\varkappa _{0}=3$ in
accordance with Feynman, which gives $A_{1}\left( 1,0\right)
=\frac{\pi ^{2}}{108}$. Then Eq.(\ref{E_0-dl2-z}) reproduces the
Feynman result for the energy bound
\begin{equation}
\mathcal{E}_{0}=-\varpi _{0}\left( \alpha +\frac{\alpha
^{2}}{81}\right) ,\;\alpha =\frac{3\pi }{4}\left( \frac{g}{\varpi
_{0}}\right) ^{2}\left( \frac{\varpi _{0}}{W}\right)
^{\frac{1}{2}},  \label{E_0deq1zeq0}
\end{equation}
while Eq.(\ref{fluct-size-dl2-z}) yields the fluctuon (polaron)
size parameter in terms of Feynman's $\alpha $ constant
\begin{equation}
l_{0}\simeq \frac{1}{6\sqrt{6}}\frac{\hbar }{\sqrt{m\varpi
_{0}}}\left( \frac{\alpha }{81}\right) ^{-1}.  \label{l_0deq1zeq0}
\end{equation}
These results have a sense upon satisfaction of
Eq.(\ref{cond1zl2}), which now reads
\begin{equation}
\frac{\left| A_{2}\left( 1,0,3\right) \right| }{3}\left( \frac{g}{\varpi _{0}%
}\right) ^{2}\left( \frac{\varpi _{0}}{W}\right) ^{\frac{1}{2}}=8\left( 7%
\sqrt{7}-18\right) \frac{\alpha }{81} \ll 1.  \label{wcdeq1zeq0}
\end{equation}

\paragraph*{Example: $d=1$, $z=1$}

This case corresponds to the interaction with acoustic-like
critical mode. Now, one should maximize the function
\begin{equation}
\overline{G}\left( 1,1,\varkappa \right) =\frac{4}{\varkappa }\left( \frac{%
3\varkappa -1}{\varkappa }\frac{\arctan \sqrt{4\varkappa -1}}{\sqrt{%
4\varkappa -1}}+\frac{\varkappa -1}{2\varkappa }\ln \varkappa
-1\right) ^{2}. \label{fun-Gbdeq1zeq1}
\end{equation}
We find numerically that the unique maximum point is $\varkappa
_{0}\simeq 3.81$ and $A_{1}\left( 1,1\right) \simeq 0.208$. Then,
Eqs.(\ref{E_0-deq2-z}) and (\ref{fluct-size-dl2-z}) yield
\begin{equation}
\mathcal{E}_{0}\simeq -\frac{3}{2}\frac{g^{2}}{W}\left[ \ln \left( \frac{W}{%
\varpi _{0}}\right) +0.104\left( \frac{g}{\varpi _{0}}\right)
^{2}\right] \label{E_0deq1zeq1}
\end{equation}
and
\begin{equation}
l_{0}K_{\max }\simeq 2.\,19\frac{W\varpi _{0}}{g^{2}},
\label{l_0deq1zeq1}
\end{equation}
respectively. In the present case the condition for the
perturbational regime, which doesn't contain $W$ at all, reads
\begin{equation}
\frac{\left\vert A_{2}\left( 1,1,\varkappa _{0}\right) \right\vert }{%
\varkappa _{0}}\left( \frac{g}{\varpi _{0}}\right) ^{2}\simeq
0.112\left( \frac{g}{\varpi _{0}}\right) ^{2}\ll 1,
\label{wcdeq1zeq1}
\end{equation}
or $\left\vert g\right\vert \ll 3\varpi _{0}$.

To conclude this section, for $0\leq z<2$ weak coupling regime is
realized at much smaller $g$ than for $z\geq 2$. For the latter,
$g$ should fit Eq.(\ref{w-bcond}) while the characteristic
fluctuation frequency $ \varpi _{0}$ plays no role. For the
former, however, the upper bound of $ \left\vert g\right\vert
/\varpi _{0}$, is crucial.

\subsection{Strong coupling regime}

\subsubsection{General Consideration}

In strong-coupling regime the electron is heavily
``fluctuation-dressed''. Let us
make in Eq.(\ref{fin-2}) the variables replacements $y=q^{-1}x$, $%
\tau=ye^{-s} $ and $t=1-\tau$. This transforms that equation to
the following one
\begin{equation}
\mathcal{E}_{0}\left( q,\lambda\right) =\frac{D}{4}Wq\left(
1-\lambda \right)
^{2}-\frac{D}{4}\frac{g^{2}}{W}q^{\frac{d+z}{2}-1}\left\langle
M\left( q,\varpi,\lambda^{2}\right) \right\rangle  \label{fin-4}
\end{equation}
where
\begin{eqnarray}
M\left( q,\varpi,\lambda^{2}\right) =\int_{0}^{q^{-1}}\frac {y^{\frac{%
d+z}{2}-1}e^{-\left( 1-\lambda^{2}\right) y}}{\epsilon\left(
y,q,\varpi,\lambda^{2}\right) }dy \nonumber \\
+\left(1-\lambda^{2}\right)
\int_{0}^{1}dt\int_{0}^{q^{-1}}\frac{1-\left( 1-t\right)
^{\epsilon\left( y,q,\varpi,\lambda^{2}\right) }}{\epsilon\left(
y,q,\varpi,\lambda^{2}\right) }e^{-\left( 1-\lambda^{2}\right) ty}y^{\frac{%
d+z}{2}}dy,  \label{fun-M}
\end{eqnarray}
and
\begin{equation}
\epsilon\left( y,q,\varpi,\lambda^{2}\right) =\frac{\varpi}{W}q^{\frac{z}{2}%
-1}y^{\frac{z}{2}}+\lambda^{2}y.  \label{fun-epsilon}
\end{equation}

To proceed, it is important to note that the function
$\epsilon\left( y,q,\varpi ,\lambda^{2}\right) $ increases, in the
integration range over $y$, from zero to $q^{-1}\epsilon$, where
$\epsilon=\frac{\varpi}{W}+\lambda^{2}$. Thus, at $q\gg\epsilon$
Eq.(\ref{fun-M}) may be expanded in asymptotic Laurent series in
overall small $\epsilon\left( y,q,\varpi,\lambda^{2}\right) $
\begin{equation}
M\left( q,\varpi,\lambda^{2}\right)
=\sum_{p=-1}^{\infty}M_{p}\left( q,\varpi,\lambda^{2}\right) ,
\label{fun-M-asym}
\end{equation}
where
\begin{equation}
M_{p}\left( q,\varpi,\lambda^{2}\right) =\int_{0}^{q^{-1}}\left[
\epsilon\left( y,q,\varpi,\lambda^{2}\right) \right]
^{p}N_{p}\left[ \left( 1-\lambda^{2}\right) y\right]
y^{\frac{d+z}{2}-1}dy  \label{fun-M_p}
\end{equation}
with $N_{-1}\left( \xi\right) =e^{-\xi}$,
\begin{equation}
N_{p}\left( \xi\right) =\frac{\left( -1\right) ^{p}}{\left( p+1\right) !}%
\int_{0}^{1}e^{-\xi t}\left[ f\left( t\right) \right]
^{p+1}t^{p+1}\xi dt,\;p\geq0,  \label{fun-N_p}
\end{equation}
and
\begin{equation}
f\left( t\right) =-\frac{\ln\left( 1-t\right) }{t}=\sum_{k=0}^{\infty }\frac{%
t^{k}}{k+1},\,0\leq t<1.  \label{fun-f}
\end{equation}
The Taylor series representing $f\left( t\right) $ converges at
$\left[ 0,1\right) $ and so does the Taylor series for any integer
power of $f\left(
t\right) $%
\begin{equation}
\left[ f\left( t\right) \right]
^{n}=\sum_{m=0}^{\infty}a_{n,m}t^{m}. \label{f^(p+1)-series}
\end{equation}

Typically, the strong coupling-regime fluctuon binding energy
$\sim Wq$ is smaller than the fluctuation energy. Hence the
above-assumed relation between $q$ and $\epsilon$ is satisfied if
$\lambda^{2}\ll q$. Another, weaker, criterion for expanding
$M_{p}\left( q,\varpi,\lambda^{2}\right) $ in powers of
$\epsilon\left( y,q,\varpi,\lambda^{2}\right) $ is inferred on by
noting that a left vicinity of $t=1$ is the dominant range for the
integration over $t$ in Eq.(\ref{fun-M}). Hence at
$\lambda^{2}\ll1$, it is the range $y\lesssim1$ that contributes
mostly to the corresponding integral
over $y$. In this range $\epsilon\left( y,q,\varpi,\lambda^{2}\right) $ $%
\lesssim \frac{\varpi}{W}q^{\frac{z}{2}-1}+\lambda^{2}$, is small,
unconditionally for $z\geq2$, and under the condition $Wq^{1-\frac{z}{2}%
}\gg\varpi$ for $z<2$. Actually, when truncating the series of
Eq.(\ref {fun-M-asym}), either $\lambda^{2}\ll q$ or
$\lambda^{2}\ll1$ and $Wq^{1- \frac{z}{2}}\gg\varpi$ are our the
\textit{only} approximations. We should check them at the end of
our calculations.

Let us try to simplify the above-developed expansion, by picking
in it up the leading terms with respect to $\kappa =\left(
1-\lambda ^{2}\right) ^{-1}q\ll 1$, not imposing in advance any
other restriction on $q$ and $ \lambda ^{2}$. To this end let us
transform $M_{p}\left( q,\varpi ,\lambda ^{2}\right) $ as follows.
For $M_{-1}\left( q,\varpi ,\lambda ^{2}\right) $, we obtain
directly
\begin{equation}
q^{\frac{d+z}{2}-1}M_{-1}\left( q,\varpi ,\lambda ^{2}\right) =\kappa ^{%
\frac{d}{2}}\int_{0}^{\kappa ^{-1}}\frac{u^{\frac{d}{2}-1}e^{-u}}{%
\frac{\varpi }{W}+\lambda ^{2}\left( \kappa u\right)
^{\frac{2-z}{2}}}du, \label{fun-M_(-1)}
\end{equation}
Further, using in Eq.(\ref{fun-M_p}) the Newton's binom, we arrive
at the identical but more convenient representation
\begin{equation}
q^{\frac{d+z}{2}-1}M_{p}\left( q,\varpi ,\lambda ^{2}\right)
=\left(
1-\lambda ^{2}\right) ^{-p-1}\sum_{k=0}^{p}C_{p}^{k}\left( \frac{\varpi }{W}%
\right) ^{k}\lambda ^{2\left( p-k\right) }m_{p+1,k+1}\left( \kappa
\right) \label{fun-M_p-sum}
\end{equation}
where
\begin{equation}
m_{n,l}\left( \kappa \right) =\frac{\left( -1\right) ^{n-1}}{n!}%
\int_{0}^{\kappa ^{-1}}\frac{\gamma \left( \frac{1}{2}%
d_{l}+n+2,u\right) }{u^{\frac{1}{2}d_{l}+1}}\left[ f\left( \kappa u\right) %
\right] ^{n}du,\;d_{l}=d+\left( z-2\right) l  \label{fun-m_(n,l)}
\end{equation}
($d_{1}=$ $d^{\ast }$ which has been introduced in Section 2 for
the case of $z\geq 2$) and
\begin{equation}
\gamma \left( b,x\right) =\int_{0}^{x}t^{b-1}e^{-t}dt,\;b>0
\label{incgamma}
\end{equation}
is the incomplete gamma-function. The integral in
Eq.(\ref{fun-M_(-1)}) at $ z\leq 2$ converges if $d>0$
irrespective of $\lambda $, while at $z>2$ this is so if $\lambda
=0$ strictly. For $\lambda \neq 0$, even small, the convergence
condition at $z>2$ reads $d_{1}>0$. These restictions upon the
critical indexes are the same as in the weak coupling regime.

The value of $M_{-1}\left( q,\varpi ,0\right) $ is independent of
$z$, and given by
\begin{equation}
q^{\frac{d+z}{2}-1}M_{-1}\left( q,\varpi ,0\right)
=\frac{W}{\varpi }\gamma \left( \frac{d}{2},\kappa ^{-1}\right)
\kappa ^{\frac{d}{2}}. \label{fun-M_-1lameq0}
\end{equation}
However, estimating $M_{-1}\left( q,\varpi ,\lambda ^{2}\right) $
at $\lambda ^{2}\neq 0$, except for the case $z=2$ where the
factor $\left( \frac{\varpi }{W}+\lambda ^{2}\right) ^{-1}$
plainly replaces $\frac{W}{\varpi }$, depends crucially upon $z$.
We postpone this task to consideration of
specific cases. At the same time, asymptotic series in $\kappa $ for $%
M_{p}\left( q,\varpi ,\lambda ^{2}\right) $ with $p\geq 0$ can be
obtained by an independent of $z$ trick.

Substituting the series of Eq.(\ref{f^(p+1)-series}) into
Eq.(\ref{fun-m_(n,l)}), integrating by parts and using the
well-known asymptotic
\begin{equation*}
\gamma \left( b,x\right) =\Gamma \left( b\right) +O\left(
x^{1-b}e^{-x}\right) ,\;x\gg 1,
\end{equation*}
we obtain with an exponential accuracy
\begin{align}
m_{n,l}\left( \kappa \right) & =\frac{\left( -1\right)
^{n-1}}{n!}\left[
\Gamma \left( \frac{1}{2}d_{l}+n+1\right) b_{n,l}\kappa ^{\frac{1}{2}%
d_{l}}-\sum_{m\neq m_{l}}^{\infty }\frac{\left( m+n\right) !a_{n,m}}{m-\frac{%
1}{2}d_{l}}\kappa ^{m}\right.  \notag \\
+& \left. c_{n,l}\left( m_{l}+n\right) !\left( \ln \kappa
^{-1}-\psi \left( m_{l}+n\right) -\frac{1}{m_{l}+n}\right) \kappa
^{m_{l}}\right] , \label{m_(n,l)-asym}
\end{align}
where $c_{n,l}=a_{n,m_{l}}$, $m_{l}$ being an integer, if any,
satisfying the condition $2m_{l}=d_{l}$, and otherwise
$c_{n,l}=0$,
\begin{equation}
b_{n,l}=\sum_{m\neq m_{l}}^{\infty
}\frac{a_{n,m}}{m-\frac{1}{2}d_{l}}, \label{b_(n,l)}
\end{equation}
and $\psi(x)$ is the logarithmic derivative of the gamma-function.
At $0<d<2$ no $m_{l}$ emerges if $z=2$, the only $m_{1}=0$ may
appear, if $z<2$ (e.g. for $z=d=1$), and if $z>2$ an infininte
number of $m_{l}\geq 1$ may exist for some $d$. It is worth noting
that at $z\leq 2$ and $d_{1}\neq 0$
\begin{equation}
b_{n,l}=\int_{0}^{1}\frac{\left[ f\left( t\right) \right] ^{n}-1}{t^{%
\frac{1}{2}d_{l}+1}}dt-\frac{2}{d_{l}}.  \label{b_(n,l)-integral}
\end{equation}

Only the cases with $m_{l}=0,1$ may be important since the
$O\left( \kappa ^{m}\ln \kappa ^{-1}\right) $ terms with $m>1$ are
small compared to the kinetic-energy term in Eq.(\ref{fin-4}). By
the same reason, of the series in integer powers of $\kappa $ in
Eqs.(\ref{m_(n,l)-asym}) we retain only the $O\left( 1\right)$
term that exists unless $m_{l}=0$. The thus approximated
Eq.(\ref{fin-4}), after performing some interim summations over
$p$ and neglecting purely non-adiabatic corrections $O\left(
\left( \frac{\varpi }{W} \right) ^{k}\right)$, becomes
\begin{align}
\mathcal{E}_{0}\left( q,\lambda \right) = & \frac{D}{4}Wq\left(
1-\lambda \right) ^{2}-\frac{D}{4}g^{2}\kappa
^{\frac{1}{2}d}\left\langle \varpi ^{-1}\Pi \left( \frac{\varpi
}{W}\kappa ^{\frac{z-2}{2}},\lambda ^{2}\right) \right\rangle -
\notag \\
& \frac{D}{4}\frac{g^{2}}{W}\Lambda \left( q,\lambda ^{2}\right)
+\Delta ,  \label{E_0sc-asym}
\end{align}
where
\begin{equation}
\Delta =-\frac{D}{2d_1}\frac{g^{2}}{W}\left( 1-\delta
_{d_{1},0}\right) \label{Delta}
\end{equation}
is an energy shift, independent of the variational parameters,
\begin{align}
\Pi \left( x,\lambda ^{2}\right) & =\int_{0}^{\infty }\frac{u^{\frac{d%
}{2}-1}e^{-u}}{1+\lambda ^{2}x^{-1}u^{\frac{2-z}{2}}}du+  \notag \\
& \sum_{n=1}^{\infty }\sum_{l=1}^{n}\frac{\left( -1\right) ^{n-1}}{n!}%
C_{n-1}^{l-1}\Gamma \left( \frac{1}{2}d_{l}+n+1\right)
b_{n,l}\left( 1-\lambda ^{2}\right) ^{-n}\lambda ^{2\left(
n-l\right) }x^{l} \label{fun-Pi}
\end{align}
and
\begin{equation}
\Lambda \left( q,\lambda ^{2}\right) =\delta _{d_{1},0}\left[ \ln \frac{1}{q}%
+\gamma +\frac{\ln \left( 1-\lambda ^{2}\right) }{\lambda
^{2}}\right]
+\delta _{d_{1},2}\left( 1-\lambda ^{2}\right) q\left( \ln \frac{1}{q}%
+\gamma -\frac{3}{2}\right)   \label{fun-Lambda}
\end{equation}
with $\gamma$ being the Euler constant. In Eq.(\ref {fun-Lambda}),
the first and the second term do not emerge at $z\geq 2$ (where
$d_{1}>0$ necessarily) and at $z<2$, respectively. In all cases
where $d_{1}>0$, $\Delta =-\mathcal{E}_{B}$, the band-edge shift
in the lowest-order Born approximation.

Further analysis on the base of Eqs.(\ref{E_0sc-asym}) -
(\ref{fun-Lambda}) depends crucially on whether $z\geq 2$ or
$z<2$. We consider these cases separately, detaching $z=2$. The
peculiarity of the latter case allows us to calculate $\Pi \left(
x,\lambda ^{2}\right) $ in a closed form and, that is not feasible
in other cases, to ultimately explore an impact of the spectral
weight $Q\left( \varpi \right) $ on the fluctuon formation.

\subsubsection{The cases with $z=2$}

For $z=2$, $d_{l}=d$ and $b_{n,l}=b_{n,n}$, so that $\Lambda
\left( q,\lambda ^{2}\right) \equiv 0$ and Eq.(\ref{fun-Pi})
greatly simplifies. The answer reads
\begin{equation}
\mathcal{E}_{0}\left( q,\lambda \right) \simeq \frac{D}{4}W\left(
1-\lambda \right) ^{2}q-\frac{D}{4}\frac{g^{2}}{W}\Gamma \left(
\frac{1}{2}d\right)
\left( 1-\lambda ^{2}\right) R_{d}\left( \lambda ^{2}\right) q^{\frac{1}{2}%
d}-\mathcal{E}_{B},  \label{fin-5}
\end{equation}
where
\begin{equation}
R_{d}\left( \lambda ^{2}\right) =\left\langle \epsilon ^{-1}\left\{ 1+\frac{d%
}{2}\int_{0}^{1}\frac{1-\left[ 1+\epsilon h\left( t\right) \right] ^{-%
\frac{1}{2}d-1}}{t^{\frac{1}{2}d+1}}dt\right\} \right\rangle .
\label{fun-R_d}
\end{equation}
and $h(t)=f(t)-1$. Formally Eq.(\ref{fin-5}) matches the case of
$\lambda =1$, as the fluctuon binding energy obtained vanishes at
$\lambda =1$, that concords with exact Eq.(\ref{fun-M}). However,
that point is likely isolated since in essential weak-coupling
regime, i.e. at $0<1-\lambda \ll 1$, the condition at $\kappa \ll
1$ may break down.

Minimization Eq.(\ref{fin-5}) first in $q$ and next in $\lambda $,
we find the optimal $q$ value
\begin{equation}
q_{0}=\left[ \Gamma \left( \frac{d}{2}+1\right) \left(
\frac{g}{W}\right)
^{2}R_{d}\left( \lambda _{0}^{2}\right) \left( \frac{1+\lambda _{0}}{%
1-\lambda _{0}}\right) \right] ^{\frac{2}{2-d}}
\label{q_0sc-zeq2}
\end{equation}
as well as the bound energy
\begin{equation}
\mathcal{E}_{0}\simeq -\frac{D}{4}\left( \frac{2}{d}-1\right)
\left[ \Gamma
\left( \frac{d}{2}+1\right) \left( \frac{g}{W}\right) ^{2}P_{d}\left( \frac{%
\varpi }{W},\lambda _{0}\right) \right]
^{\frac{2}{2-d}}W-\mathcal{E}_{B}, \label{E_0sc-zeq2}
\end{equation}
where
\begin{equation}
P_{d}\left( \lambda \right) =\left( 1+\lambda \right) \left(
1-\lambda \right) ^{1-d}R_{d}\left( \lambda ^{2}\right)
\label{fun-P_d}
\end{equation}
and $\lambda _{0}$ is the maximum point of the function
$P_{d}\left( \lambda \right) $. For $d\neq 1$,
Eq.(\ref{E_0sc-zeq2}) presents a \emph{singular} perturbation
expansion in coupling constant. When $\lambda _{0}$ corresponds to
an extremum, it satisfies the equation
\begin{equation}
2\lambda \frac{R_{d}^{\prime }\left( \lambda ^{2}\right)
}{R_{d}\left( \lambda ^{2}\right) }+\frac{d-\left( 2-d\right)
\lambda }{1-\lambda ^{2}}=0, \label{lambda_0-eq}
\end{equation}
otherwise $\lambda _{0}=0$. For the latter case,
\begin{equation*}
P_{d}\left( \lambda _{0}\right) =R_{d}\left( \lambda
_{0}^{2}\right) =\frac{W}{\varpi_0}+O\left( 1\right),
\end{equation*}
where $\varpi_0 = \left\langle \varpi^{-1}\right\rangle^{-1}$,
which attains, to within $O\left( 1\right) $ terms, largest of all
possible values of those functions. Note that $\lim_{\varpi
\rightarrow 0}Q\left( \varpi \right) =0$, so it is likely that
$\left\langle \varpi ^{-1}\right\rangle <\infty $.

Let us search a solution $\lambda _{0}$ to Eq.(\ref{lambda_0-eq}),
in the vicinity of $\lambda =0$. Assuming that also $\left\langle
\varpi ^{-2}\right\rangle <\infty $, we have in the leading
approximation
\begin{equation}
R_{d}\left( \lambda ^{2}\right) \simeq W\left\langle \varpi
^{-1}\right\rangle -\lambda ^{2}W^{2}\left\langle \varpi
^{-2}\right\rangle , \label{R_d-approx}
\end{equation}
which yields for the sought solution
\begin{equation}
\lambda _{0}\simeq \frac{d}{2}\frac{\varpi _{1}}{W},\;\varpi _{1}=\frac{%
\left\langle \varpi ^{-1}\right\rangle }{\left\langle \varpi
^{-2}\right\rangle }.  \label{lambda_opt}
\end{equation}
For ``rigid'' $Q\left( \varpi \right) $, i.e. zeroing below some
finite $ \varpi $, the above-exploited assumption $\left\langle
\varpi ^{-2}\right\rangle <\infty $ holds automatically. Consider
now ``soft'' $ Q\left( \varpi \right) $, for which $\left\langle
\varpi ^{-2}\right\rangle =\infty $, but $\left\langle \varpi
^{-1-\sigma }\right\rangle <\infty $ with some $0<\sigma <1$.
Scaling the behavior of $Q\left( \varpi \right) $ at $\varpi
\rightarrow 0^{+}$ by
\begin{equation*}
Q\left( \varpi \right) \sim b_{\sigma }\left\langle \varpi
^{-1-\sigma }\right\rangle \frac{\sin \left( \pi \sigma \right)
}{\pi \sigma }\varpi ^{\sigma },\;b_{\sigma }=\mbox{const},
\end{equation*}
we obtain the solution to Eq.(\ref{lambda_0-eq}) at $1>\sigma
>1/2$
\begin{equation}
\lambda _{0}\simeq \left( \frac{d}{2b_{\sigma }}\right) ^{\frac{1}{2\sigma -1%
}}\left( \frac{\varpi _{\sigma }}{W}\right) ^{\frac{\sigma }{2\sigma -1}%
},\;\varpi _{\sigma }=\left( \frac{\left\langle \varpi ^{-1}\right\rangle }{%
\left\langle \varpi ^{-1-\sigma }\right\rangle }\right)
^{\frac{1}{\sigma }}. \label{lambda_0-soft}
\end{equation}
If $0<\sigma \leq 1/2$, $\lambda _{0}$ remains zero. Since $\frac{\sigma }{%
2\sigma -1}>1$ at $1>\sigma >1/2$, the non-adiabatic corrections
resulting
from $\lambda _{0}\sim \left( \frac{\varpi }{W}\right) ^{\frac{\sigma }{%
2\sigma -1}}$ are even smaller than those $\sim \frac{\varpi }{W}$
resulting from the integral term in Eq.(\ref{fun-R_d}).

Thus, as far as small $\lambda _{0}$ is concerned, either $\lambda
_{0}^{2}=o\left( \frac{\varpi }{W}\right) $ or $\lambda _{0}=0$
for all admissible $Q\left( \varpi \right) $. Neglecting the
postleading non-adiabatic corrections, from Eqs.(\ref{q_0sc-zeq2})
and (\ref{E_0sc-zeq2}) we arrive at
\begin{equation}
q_{0}=\left[ \Gamma \left( \frac{d}{2}+1\right) \frac{g^{2}}{W\varpi _{0}}%
\right] ^{\frac{2}{2-d}},\;\mathcal{E}_{0}=-\frac{D}{4}\left( \frac{2}{d}%
-1\right) Wq_{0}-\mathcal{E}_{B}.  \label{q_0, E_0sc}
\end{equation}
Requiring $q_{0}\ll 1$, one gets the criterion of applicability of
the continuum approximation
\begin{equation}
\Gamma \left( \frac{d}{2}+1\right) \frac{g^{2}}{W\varpi _{0}}<1.
\label{w-bsccond}
\end{equation}
Under this condition, the self-trapping term $\propto \left( \frac{g^{2}}{%
\varpi _{0}W}\right) ^{\frac{2}{2-d}}$ in $\mathcal{E}_{0}$ may be
both smaller and larger than $\mathcal{E}_{B}$. The latter
situation occurs if coupling is strong enough to satisfy
\begin{equation}
\Gamma \left( \frac{d}{2}+1\right) \frac{g^{2}}{W\varpi _{0}}>\left[ \frac{2%
}{d}\frac{2}{2-d}\frac{1}{\Gamma \left( \frac{1}{2}d\right) }\right] ^{\frac{%
2-d}{d}}\left( \frac{\varpi _{0}}{W}\right) ^{\frac{2-d}{d}}.
\label{sccon-zeq2}
\end{equation}
Even though $-\mathcal{E}_{B}$ dominates $\mathcal{E}_{0}$, the
self-trapping term yet lowers $\mathcal{E}_{0}$ more than does the
correction $\propto \left( \frac{g^{2}}{W^{2}}\right) ^{2}$ in
weak coupling regime.

Consider now the singular case $d=2$, for which $m_{1}=1$ and
\begin{equation*}
b_{n}=\int_{0}^{1}\frac{\left[ f\left( t\right) \right] ^{n}-1-\frac{1%
}{2}nt}{t^{2}}dt-1.
\end{equation*}
Here we obtain from Eq.(\ref{E_0sc-asym})
\begin{align}
\mathcal{E}_{0}\left( q,\lambda \right) = & \frac{D}{4}q\left[
W\left( 1-\lambda \right) ^{2}-\frac{g^{2}}{W}\left( 1-\lambda
^{2}\right) R_{2}\left( \lambda ^{2}\right) \right] + \notag
\\
& \frac{D}{4}\frac{g^{2}}{W}\left( 1-\lambda ^{2}\right) q
\ln\frac{q}{e^{\frac{3}{2}-\gamma }}-\mathcal{E}_{B}
\end{align}
where
\begin{equation*}
R_{2}\left( \lambda ^{2}\right) =\left\langle \epsilon
^{-1}+\int_{0}^{1}\allowbreak \left[ \frac{h\left( t\right)
}{\left(
1+\epsilon h\left( t\right) \right) ^{2}}+\frac{h\left( t\right) }{%
1+\epsilon h\left( t\right) }-t\right]
\frac{dt}{t^{2}}\right\rangle
\end{equation*}
This expression is easily optimized first over $q$ and afterwards
over $\lambda $ to yield
\begin{equation}
q_{0}=e^{\frac{1}{2}-\gamma +S\left( \lambda _{0}\right) },\;\mathcal{E}%
_{0}=-\left( 1+q_{0}\right) \mathcal{E}_{B},  \label{q_0,
E_0deqzeq2}
\end{equation}
where
\begin{equation*}
\;S\left( \lambda \right) =R_{2}\left( \lambda ^{2}\right) -\left( \frac{W}{g%
}\right) ^{2}\frac{1-\lambda }{1+\lambda }.
\end{equation*}
and $\lambda _{0}$ is the maximum point of the function $S\left(
\lambda \right) $. Searching again $\lambda _{0}\ll 1$, we obtain
\begin{equation}
\lambda _{0}\simeq \frac{1}{g^{2}\left\langle \varpi ^{-2}\right\rangle }%
,\;q_{0}=e^{2-\gamma -\frac{W}{\varpi _{0}}\left( \frac{W\varpi _{0}}{g^{2}}%
-1\right) }.  \label{q_0, lam0-small}
\end{equation}
$\allowbreak $It is seen that for $\lambda _{0}\ll 1$ and
$q_{0}\ll 1$, the inequality $g^{2}\left\langle \varpi
^{-2}\right\rangle \gg 1$ and Eq.(\ref {w-bsccond}) with $d=2$,
respectively, should hold. $\mathcal{E}_{0}$ given by
Eqs.(\ref{q_0, E_0deqzeq2}), (\ref{q_0, lam0-small}) is much above
that obtained in weak-coupling regime (see case $d^{\ast }=2$ in
previous subsection) for typically $\frac{W\varpi
_{0}}{g^{2}}-1=O\left( 1\right) $, though if $\frac{W\varpi
_{0}}{g^{2}}-1=O\left( \frac{\varpi _{0}}{W}\right) $ the former
may gain.

But what happens if Eq.(\ref{w-bsccond}) doesn't hold ? The answer
is easy for $2\geq d>1$ - in this case weak coupling regime may
realize. For $d\leq 1 $, however, the question cannot be answered
within the present framework, as numerical study reveals no any
maximum of $P_{d}\left( \lambda \right) $ other than that in a
close vicinity of $\lambda =0$.

\subsubsection{The cases with $z\neq 2$}

Using the experience with $z=2$, in what follows we restrict
ourselves to small $\lambda $, and assume $\left\langle \varpi
^{-1-s }\right\rangle <\infty $, where $0\leq s \leq 1$
throughout. The integral part of $\Pi \left( x,\lambda ^{2}\right)
$ possesses small-$\lambda$ expansion at $\lambda ^{2}\ll x$,
which we force to hold. Further, we have $x\ll 1$ unconditionally
if $z>2$. For $z<2$ we force holding $x\ll 1$ anymore. At the end,
we check those conditions both to hold. With such prerequisites,
up to the first-order terms inclusive, we obtain
\begin{equation}
\mathcal{E}_{0}\left( q,\lambda \right) =\mathcal{E}_{0}\left( q,0\right) -%
\frac{D}{2}Wq\lambda +\frac{D}{4}W\Gamma \left(
1+\frac{d-z}{2}\right) g^{2}\left\langle \varpi ^{-2}\right\rangle
q^{1+\frac{d-z}{2}}\lambda ^{2}, \label{E_0dgz-2}
\end{equation}
at $d>z-2$ and
\begin{equation}
\mathcal{E}_{0}\left( q,\lambda \right) =\mathcal{E}_{0}\left( q,0\right) -%
\frac{D}{2}Wq\lambda +\frac{D}{2d}\frac{\pi \sigma }{\sin \pi \sigma }\frac{%
\left[ q\left( 0\right) \right] ^{\frac{2-d}{2}}}{\Gamma \left( 1+\frac{d}{2}%
\right) }\left( \frac{W}{\varpi _{\sigma }}\right) ^{\sigma
}\lambda ^{2\sigma },\, \label{E_0dlz-2}
\end{equation}
at $d>z-2$, where $\sigma =\frac{d}{z-2}$. Here
\begin{align}
\mathcal{E}_{0}\left( q,0\right) = &
\frac{D}{4}Wq-\frac{D}{2d}\left[ q\left( 0\right) \right]
^{\frac{2-d}{2}}q^{\frac{1}{2}d}-\frac{D}{4}\left[ q\left(
0\right) \right] ^{\frac{2-d}{2}}\frac{\varpi _{0}}{W}\frac{\Gamma
\left( \frac{z+d}{2}\right) }{\Gamma \left( \frac{d}{2}+1\right)
}bq^{\frac{d+z}{2} -1} \notag \\
& +\Delta -\frac{D}{4}\frac{g^{2}}{W}\Lambda \left( q,0\right),
\label{E_0q_0}
\end{align}
$q\left( 0\right)$ is $q_{0}$ obtained with $\lambda =0$, i.e.
given by Eq.(\ref{q_0, E_0sc}) and
\begin{equation}
b=b_{1,1}=-\frac{1}{z+d}\left\{
\begin{array}{r}
2\psi \left( 1-\frac{z+d}{2}\right) +2\gamma ,\;\,z+d\neq 4 \\
1,\;z+d=4 \label{b}
\end{array}
\right.
\end{equation}

Let $d_{1}\neq 0,1$, i.e. $\Lambda \left( q,0\right) =0$. For
$d>z-2$, the minimization equation for $\lambda $ is solved to
give
\begin{equation}
\lambda \left( q\right)
=\frac{q^{\frac{z-d}{2}}}{g^{2}\left\langle \varpi
^{-2}\right\rangle \Gamma \left( 1+\frac{d-z}{2}\right) }.
\label{lam-q}
\end{equation}
Then the minimization equation for $q$ is well solved by
iterations in small adiabatic parameter, to yield for the
variational parameters
\begin{eqnarray}
q_{0} &\simeq &q\left( 0\right) -\frac{2}{2-d}\frac{\varpi
}{W}\left[ q\left( 0\right) \right] ^{\frac{z}{2}}
\label{q_0omega} \\
\lambda _{0} &\simeq &\frac{\Gamma \left( 1+\frac{d}{2}\right)
}{\Gamma \left( 1+\frac{d-z}{2}\right) }\frac{\varpi
_{1}}{W}\left[ q\left( 0\right) \right] ^{\frac{z}{2}-1}
\label{lamomega}
\end{eqnarray}
where {\it ad hoc} $\varpi$ is defined by
\begin{equation}
\varpi = \frac{\left(1-\frac{z+d}{2}\right)\Gamma
\left(\frac{z+d}{2}\right) b}{\Gamma \left(
1+\frac{d}{2}\right)}\varpi_{0} +
\frac{\left(\frac{d-z}{2}-1\right)\Gamma \left( 1+\frac{d}{2}
\right) }{\Gamma \left( 1+\frac{d-z}{2}\right) }\varpi _{1}
\label{ad hoc}
\end{equation}

For $d<z-2$ that may realize only at $z>2$, we find the optimal
$\lambda$ at a given $q$ to equal
\begin{equation}
\lambda \left( q\right) =\left[ \frac{z-2}{2}\frac{\sin \pi \sigma
}{\pi
\sigma }\Gamma \left( 1+\frac{d}{2}\right) \right] ^{\frac{1}{2\sigma -1}%
}\left( \frac{\varpi _{\sigma }}{W}\right) ^{\frac{\sigma }{2\sigma -1}}%
\left[ q\left( 0\right) \right] ^{-\frac{2-d}{2}\frac{1}{2\sigma -1}}q^{%
\frac{1}{2\sigma -1}}, \label{lamqdlz-2}
\end{equation}
which provides a minimum if $2\sigma >1$(i.e. $d>\frac{z-2}{2}$),
otherwise we should put $\lambda =0$. Just as above, the equation
for optimum $q$ at the present conditions is solved by iterations,
which results in
\begin{eqnarray}
q &\simeq &q\left( 0\right) -\frac{2}{2-d}\frac{\varpi }{W}\left[
q\left(0\right) \right] ^{\frac{z}{2}} \label{q0-omega1} \\
\lambda _{0} &\simeq &\left[ \frac{z-2}{2}\frac{\sin \pi \sigma
}{\pi \sigma }\Gamma \left( 1+\frac{d}{2}\right) \right]
^{\frac{1}{2\sigma -1}}\left(
\frac{\varpi _{\sigma }}{W}\left[ q\left( 0\right) \right] ^{\frac{z}{2}%
-1}\right) ^{\frac{\sigma }{2\sigma -1}}, \label{lam0-omega1}
\end{eqnarray}
where here $\varpi$ denotes only the first term in the expression
given by Eq.(\ref{ad hoc}). To check all necessary conditions, we
consider below the cases with $z>2$ and $z<2$ separately.

\paragraph*{Subcase $z>2$}

For $z>2$, we have from Eqs.(\ref{q_0omega})-(\ref{lam0-omega1})
$q_{0}=q\left( 0\right) +o\left( \frac{\varpi }{W}\right) $ and
$\lambda _{0}=o\left( \frac{\varpi }{W}\right) $. Both $\lambda
_{0}\ll 1$ and $\lambda _{0}^{2}\ll \frac{\varpi _{\sigma
}}{W}q_{0}^{\frac{z }{2}-1}$ are satisfied automatically. So the
corrections to formula for $ \mathcal{E}_{0}$ as given above for
the cases with $z=2$ are much smaller than $\varpi$ and even not
worth to be considered anymore. There remain the same conditions,
given by Eqs.(\ref{w-bsccond}) and (\ref{sccon-zeq2}), as with
$z=2$.

\paragraph*{Subcase $z<2$:}

For $z<2$ and $d+z-2\neq 0$, using Eq.(\ref{q_0omega}) we have for
original parameter $v=qW$ up to the first order corrections
\begin{equation}
v_{0}=q_{0}W\simeq W\left[ \Gamma \left( \frac{d}{2}+1\right) \frac{g^{2}}{%
W\varpi _{0}}\right] ^{\frac{2}{2-d}}-\frac{2}{2-d}\left[ \Gamma
\left(
\frac{d}{2}+1\right) \frac{g^{2}}{W\varpi _{0}}\right] ^{\frac{z}{2-d}%
}\varpi ,\;  \label{v_0-zl2}
\end{equation}
where, as introduced above,
\begin{equation}
\varpi =2\frac{\left( 1-\frac{z+d}{2}\right) \Gamma \left( \frac{z+d}{2}%
\right) b}{\Gamma \left( 1+\frac{d}{2}\right) }\varpi
_{0}+\frac{\left(
\frac{d-z}{2}-1\right) \Gamma \left( 1+\frac{d}{2}\right) }{\Gamma \left( 1+%
\frac{d-z}{2}\right) }\varpi _{1},  \label{omega-zl2}
\end{equation}
for the parameter $\lambda$
\begin{equation}
\lambda _{0}\simeq \frac{\Gamma \left( 1+\frac{d-z}{2}\right)
}{\Gamma
\left( 1+\frac{d}{2}\right) }\left[ \Gamma \left( \frac{d}{2}+1\right) \frac{%
g^{2}}{W\varpi _{0}}\right] ^{-\frac{2-z}{2-d}}\frac{\varpi
_{1}}{W} \label{lam_0-zl2}
\end{equation}
and for the fluctuon energy
\begin{equation}
\mathcal{E}_{0}=-\frac{D}{4}\left( \frac{2}{d}-1\right) W\left[
\Gamma
\left( \frac{d}{2}+1\right) \frac{g^{2}}{W\varpi _{0}}\right] ^{\frac{2}{2-d}%
}+\Delta -\frac{D}{4}\left[ \Gamma \left( \frac{d}{2}+1\right) \frac{g^{2}}{%
W\varpi _{0}}\right] ^{\frac{z}{2-d}}\varpi_{01}, \label{E_0zl22}
\end{equation}
where
\begin{equation}
\varpi_{01}=\frac{\Gamma \left( \frac{z+d}{2}\right) b}{\Gamma
\left(
1+\frac{d}{2}\right) }\varpi _{0}+\frac{\Gamma \left( 1+\frac{d}{2}\right) }{%
\Gamma \left( 1+\frac{d-z}{2}\right) }\varpi _{1}  \label{omegapr}
\end{equation}

Now the check of necessary conditions is in order. If we require
that $O\left( \lambda ^{2}x^{-1}\right)$ terms should be small on
average, we get
\begin{equation}
\Gamma \left( \frac{d}{2}+1\right) \frac{g^{2}}{\varpi _{0}W}\gg
\left[ |b| \frac{\Gamma \left( 1+\frac{d-z}{2}\right) }{\Gamma
\left( 1+\frac{d}{2} \right) }\left( \frac{\varpi _{0}}{W}\right)
\right] ^{\frac{2-d}{2-z}}. \label{sccond1-zl2}
\end{equation}
The conditions that $x \ll 1$ and $\lambda _{0}\ll 1$, to within
purely numerical factor, give the same inequality as
Eq.(\ref{sccond1-zl2}).

Note that at $d+z-2<0$ the value $\Delta >0$ and has no connection
to $\mathcal{E}_{B}$. In these subcases, Eq.(\ref{sccond1-zl2})
proves much stronger than that of Eq.(\ref{sccon-zeq2}) that leads
to total domination of the self-trapping energy term over
$\Delta$. Moreover to within the present approximation, $\Delta$
is much smaller even than the $O\left( \varpi \right) $ correction
in $\mathcal{E}_{0}$. As an example of such a case, consider again
Feynman polaron ($D=3$, $d=1$, $z=0$). From Eq.(\ref{b}) we have
$b=4\ln 2$ and from Eqs.(\ref{v_0-zl2}) - (\ref {lam_0-zl2}) we
obtain, in terms of Feynman's $\alpha$, for the original
variational parameters $v=Wq$ and $ w=\lambda v$
\begin{equation*}
v_{0}\simeq \left( \frac{4\alpha ^{2}}{9\pi }+1-8\ln 2\right)
\varpi _{0},\;w\simeq \varpi _{0}
\end{equation*}
and for the energy
\begin{equation*}
\mathcal{E}_{0}=-\left( \frac{\alpha ^{2}}{3\pi }+6\ln
2+\frac{3}{4}\right) \varpi _{0},
\end{equation*}
These results are valid upon the conditions
\begin{equation}
\frac{3}{2}\sqrt{\frac{\pi W}{\varpi _{0}}}>\alpha \gg
\frac{3}{2}\sqrt{2\pi \ln 2}\simeq 3.
\end{equation}
The left-hand side inequality (particular case of
Eq.(\ref{w-bsccond})) doesn't appear in Feynman theory, since
there $W=\infty$.

At the end, explore singular cases, with $m_{1}=0$, i.e. $d+z=2$.
Using Eqs.(\ref{E_0sc-asym}) - (\ref{fun-Lambda}) we obtain
\begin{align}
\mathcal{E}_{0}\left( q,0\right) = & \frac{D}{4}Wq-\frac{D}{2}Wq\lambda +\frac{D%
}{4}W\Gamma \left( d\right) g^{2}\left\langle \varpi
^{-2}\right\rangle q^{d}\lambda ^{2}- \notag \\
& \frac{D}{2d}\left[ q\left( 0\right) \right] ^{\frac{2-d}{2}%
}q^{\frac{d}{2}}+\frac{D}{4}\frac{g^{2}}{W}\ln \left( \frac{q}{e^{1-\gamma }}%
\right) . \label{E_0d+z2}
\end{align}
As above, the minimization equation for $\lambda$ is solved
exactly
\begin{equation}
\lambda \left( q\right) =\frac{q^{1-d}}{g^{2}\left\langle \varpi
^{-2}\right\rangle \Gamma \left( d\right) }, \label{lamqd+z2}
\end{equation}
while that for $q=q\left( 0\right) y$, being
\begin{equation*}
y=\left\{ 1-\frac{2-d}{\Gamma \left( d\right) }\frac{\varpi
_{1}}{W}\left[
q\left( 0\right) \right] ^{-\frac{d}{2}}y^{1-d}\Gamma \left( 1+\frac{d}{2}%
\right) +\frac{\varpi _{0}}{W}y^{-1}\Gamma \left(
1+\frac{d}{2}\right) \right\} ^{-\frac{2}{2-d}}
\end{equation*}
is well solved by iterations around $y=1$, to yield
\begin{equation}
q_0 \simeq q\left( 0\right) +\left\{ \left[ q\left( 0\right) \right] ^{1-\frac{d%
}{2}}\frac{2}{\Gamma \left( d\right) }\frac{\varpi _{1}}{W}-\frac{2}{2-d}%
\frac{\varpi _{0}}{W}q\left( 0\right) \right\} \Gamma \left( 1+\frac{d}{2}%
\right). \label{q01}
\end{equation}
Eq.(\ref{q01}) is valid provided that
\begin{equation*}
\Gamma \left( 1+\frac{d}{2}\right) \frac{2-d}{\Gamma \left( d\right) }\frac{%
\varpi _{1}}{W}\left[ q\left( 0\right) \right] ^{-\frac{d}{2}}\ll
1,
\end{equation*}
which means a sort of strong-coupling conditions, considered above
\begin{equation}
\Gamma \left( 1+\frac{d}{2}\right) \frac{g^{2}}{\varpi _{0}W}\gg
\left[ \frac{2-d}{\Gamma \left( d\right) }\Gamma \left(
1+\frac{d}{2}\right) \left( \frac{\varpi _{1}}{W}\right) \right]
^{\frac{2-d}{d}}. \label{sccond1}
\end{equation}
Then using Eq.(\ref{E_0d+z2}), (\ref{lamqd+z2}), and
Eq.(\ref{q01}) we obtain in the leading approximation
\begin{equation}
q_{0}\simeq \left[ \Gamma \left( \frac{d}{2}+1\right)
\frac{g^{2}}{W\varpi
_{0}}\right] ^{\frac{2}{2-d}}+\frac{2\left[ \Gamma \left( 1+\frac{d}{2}%
\right) \right] ^{2}}{\Gamma \left( d\right) }\left( \frac{g}{W}\right) ^{2}%
\frac{\varpi _{1}}{\varpi _{0}}, \label{q0}
\end{equation}
\begin{equation}
\lambda _{0}\simeq \frac{\Gamma \left( \frac{d}{2}+1\right)
}{\Gamma \left(
d\right) }\left[ \Gamma \left( \frac{d}{2}+1\right) \frac{g^{2}}{W\varpi _{0}%
}\right] ^{-\frac{d}{2-d}}\frac{\varpi _{1}}{W}, \label{lam0}
\end{equation}
and the energy
\begin{align}
\mathcal{E}_{0} =  & -\frac{D}{4}\left( \frac{2}{d}-1\right)
W\left[ \Gamma \left( \frac{d}{2}+1\right) \frac{g^{2}}{W\varpi
_{0}}\right] ^{\frac{2}{2-d} } \notag \\
& +\frac{D}{4}\frac{g^{2}}{W}\ln \left( e^{\gamma -1}\left[ \Gamma
\left( \frac{d}{2}+1\right) \frac{g^{2}}{W\varpi _{0}}\right]
^{\frac{2}{2-d} }\right) -\frac{3D}{4}\frac{g^{2}}{W}\frac{\left[
\Gamma \left( 1+\frac{d}{2} \right) \right] ^{2}}{\Gamma \left(
d\right) }\frac{\varpi _{1}}{\varpi _{0}} . \label{E00}
\end{align}

As an example of the peculiar case $d+z=2$ one may consider
$z=d=1$. Assuming for simplicity $\varpi _{1}=\varpi _{0}$ we
obtain
\begin{equation*}
q_{0}\simeq \frac{\pi }{4}\left[ \left( \frac{g}{\varpi _{0}}\right) ^{2}+2%
\right] \left( \frac{g}{W}\right) ^{2},\;\lambda _{0}\simeq \left( \frac{%
\varpi _{0}}{g}\right) ^{2},
\end{equation*}
and
\begin{equation}
\mathcal{E}_{0}=-\frac{D}{16}\pi \left[ \left( \frac{g}{\varpi
_{0}}\right)
^{2}+3\right] \frac{g^{2}}{W}+\frac{D}{2}\frac{g^{2}}{W}\ln \left( \frac{%
\sqrt{\pi e^{\gamma -1}}}{2}\frac{g^{2}}{W\varpi _{0}}\right) .
\label{scE_0-zeqdeq1}
\end{equation}
Strong-coupling condition (\ref{sccond1}) in the present example
simplifies to
\begin{equation*}
\left( \frac{g}{\varpi _{0}}\right) ^{2}\gg 1
\end{equation*}
It appears that this condition and weak-coupling condition given
by Eq.(\ref{wcdeq1zeq1}) have wide overlap, within which
Eq.(\ref{scE_0-zeqdeq1}) results in much lower $\mathcal{E}_{0}$
than Eq.(\ref{E_0deq1zeq1}). Even the absolute value of
logarithmic correction proves larger than that of Born shift
$\frac{D}{2}\frac{g^{2}}{W}\ln \left( \frac{ W}{\varpi
_{0}}\right)$. This means that the strong coupling solution is
energetically more favorable in the overlap region.

\section{The self-trapping and electron density of states at classical
critical point}

\subsection{Variational estimation for the electron free energy}

Let us consider now the self-trapping of the electron at a
\textit{classical} critical point, CCP (or second-order phase
transition) at finite temperature $T_{c}=\beta_{c}^{-1}$
(rigorously speaking, the transition can be considered as a
classical one only assuming that it is not too close to QCP at
zero temperature \cite{sachdev}). The Feynman variational approach
has been applied to this problem by us earlier
\cite{ourTMF,ourJMMM,ourDAN} (only for a particular case
$D=3,\eta=0$) but here we reconsider this (for a generic
situation) concentrating on some new points such as the behavior
of the electron density of states (DOS) and detailed comparison
with the quantum case treated above.

We start with the same general expression given by Eq.(\ref
{finest}). Typically for CCP one has $\hbar \varpi\beta_{c}\ll1$
due to well-known phenomenon of critical slowing down
\cite{halperin}. This is true provided that a typical wave vector
of the order-parameter fluctuations is small in comparison with
the reciprocal lattice vector; in our case the typical wave
vectors $K^{\ast}\simeq1/l_{0}$ (where $l_{0}$ is an optimal
fluctuon size) should be much smaller than $K_{\max}$ and
therefore, indeed, $\hbar\varpi\beta_{c}\simeq\left( K^{\ast
}/K_{\max}\right) ^{z}\ll1$ so we can use for our estimations
long-wavelength asymtotic of \textit{static} order-parameter
correlators. Due to irrelevance of the dynamics one can put it the
trial action (\ref{tract}) $w=0$. We will also use the notation
$C=\omega^{2}/2$ where $\omega$ is the frequency of the trial
oscillator; the fluctuon size is $l=\left( \hbar/2m\omega\right)
^{1/2}.$ We will be interested in the strong-coupling regime where
\begin{equation}
\beta_{c}\hbar\omega\gg1  \label{ineq_cl}
\end{equation}
Then instead of Eq.(\ref{Gaussest}) we will have for the Gaussian
case the following estimation (cf. Ref.\cite{ourTMF})
\begin{equation}
\mathcal{F}\leq\frac{D\omega}{4}-\frac{\beta g^{2}A_{D}}{2} \int
_{0}^{K_{\max}}\mathcal{K}_{2}\left( K\right) \exp\left( -
\frac{K^{2}}{2\omega}\right) K^{D-1}dK \label{Gaus_cl}
\end{equation}
where $\mathcal{K}_{2}\left( K\right)$ is the Fourier transform of
the static order-parameter correlation function with a small-$K$
expression
\begin{equation}
\mathcal{K}_{2}\left( K\right) = \left ( \frac{K_{{\rm max}}}{K}
\right ) ^{2-\eta} \label{statcorr}
\end{equation}
A numerical factor factor in the above expression is absorbed into
the coupling constant $g$. For the reasons which will be clear
below we consider $\beta$ in the partition function and, as a
consequence, in Eq.(\ref{Gaus_cl}), a running variable.
Substituting Eq.(\ref{statcorr}) into Eq.(\ref{Gaus_cl}) one
promptly finds
\begin{equation}
\mathcal{F}\leq\frac{D\omega}{4}-\frac{D\beta g^2}{4}\Gamma \left(
\frac{d}{2} \right ) \left ( \frac{\omega}{W} \right)^{d/2}
\label{Gauss_cl1}
\end{equation}
After minimization of the right-hand side of Eq.(\ref{Gauss_cl1})
we find for the optimal estimation of the electron free energy
\begin{equation}
\mathcal{F}_{0}\left( \beta,g\right) =-\frac{DW(2-d)}{4d} \left [
\Gamma \left ( \frac{d}{2}+1 \right ) \frac{\beta g^{2}}{W} \right
] ^{\frac{2}{2-d} }\equiv-BW\left( \frac{\beta g^{2}}{W}\right)
^{\frac{2}{2-d}} \label{clas_opt}
\end{equation}
Similar to Ref.\cite{ourTMF} one can show that this is an optimal
estimation provided that
\begin{equation}
\left( \beta W\right) ^{d/2}\ll\left( \beta g\right)^{2}\ll\left(
\beta W\right) ^{d}  \label{doubleineq}
\end{equation}
where the left inequality gives the criterion of the strong
coupling, or self-trapping, and the right one gives the criterion
of applicability of the Gaussian approximation. The latter is
found from the consideration of the scaling properties of
higher-order cumulants in the expansion (\ref{finest}). For
$\left( \beta g\right) ^{2}\ll\left( \beta W\right) ^{d/2}$ (weak
coupling regime) the second-order Born approximation turns out to
be optimal. For $d=1$ these results coincide with that from Ref.
\cite{ourTMF}.

Comparing the result (\ref{clas_opt}) with the ground-state energy
estimations for strong-coupling regime (\ref{q_0,
E_0sc}),(\ref{E_0zl22}) one can see that in the leading order
these expressions differ just by a natural replacement of the
temperature $\beta^{-1}$ for the classical critical point by a
typical fluctuation energy for the quantum case. However, the
physical meaning of these quantities is essentially different:
whereas for the quantum case we have derived an estimation for the
true boundary of the electron energy spectrum, for the classical
one our result is connected with the fluctuation density of states
tail which is not restricted (in the Gaussian approximation) from
below. Further we will prove this important statement.

\subsection{Electron density of states tail: Laplace transformation}

The electron partition function (\ref{partfun}) can be estimated,
due to Eq.(\ref{clas_opt}), as
\begin{equation}
Z\simeq\exp\left(
BW^{-\frac{d}{2-d}}\beta^{\frac{4-d}{2-d}}g^{\frac{4}{2-d}}\right)
, \label{Zestim}
\end{equation}
At the same time it can be rigorously expressed as a Laplace
transform of the electron DOS
\begin{equation}
N\left( E\right) =\left\langle \delta\left( E-\mathcal{H}\right)
\right\rangle _{f},  \label{DOS}
\end{equation}
namely,
\begin{equation}
Z=\int _{0}^{\infty}N\left( E\right) e^{-\beta E}dE
\label{Laplace}
\end{equation}
We can use now Eqs.(\ref{Zestim}),(\ref{Laplace}) to find the
asymtotic of the electron density of states (that is why it was
important to consider $\beta$ formally as an independent
variable). Using the saddle point method one can prove that at
large enough negative $E$
\begin{equation}
N\left( E\right) \propto \exp \left[ - \frac{1}{2}\left (
\frac{4}{4-d} \right)^{2-\frac{d}{2}} \left ( \frac{D}{d} \right
)^{\frac{d}{2}}\Gamma \left(1-\frac{d}{2} \right ) \left(
\frac{\left| E\right|}{E_0}\right)^{2-d/2} \right] \label{density}
\end{equation}
with a suitable choice of the energy scale $E_{0}$ as
\begin{equation}
E_{0}=\left( \frac{\pi D}{2\sin \frac{\pi d}{2}} \right
)^\frac{2}{4-d} g^{\frac{4}{4-d}}W^{-\frac{d}{4-d}}\label{E00}
\end{equation}
(origin of a numerical factor in the definition (\ref{E00}) will
become clear in the next Subsection). The saddle point method is
applicable if the exponential in the above formula is large, which
is connected with the left inequality in Eq.(\ref {doubleineq}).
Another restriction is obvious from the observation that the real
edge of the spectrum for the Hamiltonian (\ref{ham}) without
fluctuation dynamics equal to $E_{min}=- g \max\left|\varphi
\right|$. Therefore the asymptotic (\ref{density}) makes sense
only for $\left| E\right| \ll\left| g\right|$. Near the edge of
the spectrum the ``Gaussian'' tail (\ref{density}) transforms into
the ``Lifshitz'' one. Analyzing the scaling properties of the
higher-order cumulants one can demonstrate that at $E\rightarrow
E_{min}+0$
\begin{equation}
N\left( E\right) \varpropto\exp\left[ -\frac{const}{\left(
E-E_{min}\right) ^{d/2}}\right].  \label{lifshitz}
\end{equation}
This result has been obtained in Ref.\cite{ourDAN} for $d=1$.

\subsection{DOS tail: diagrammatic approach}

To better appreciate the above-mentioned approximations, it is
instructive to reproduce the result (\ref{density}) by another way
basing on the diagram technique \cite{mahan,migdal,doniach}. The
average Green function of the electron describing by the
Hamiltonian (\ref{ham}) with the Gaussian random static field
$\varphi\left( \mathbf{r}\right)$ is written in a closed form
\begin{align}
G(E,\mathbf{P}) & =\frac{1}{E-\mathbf{P}^{2}/2-\Sigma\left(E,\mathbf{P}\right)}  \notag \\
\Sigma\left(E,\mathbf{P}\right) & =g^{2}\Omega_{D}\int\gamma\left(
\mathbf{P-K},\mathbf{P},\mathbf{K};E\right) \mathcal{K} _{2}\left(
\mathbf{K}\right) G(E,\mathbf{P-K})\frac{d^{D}K}{\left(
2\pi\right)^{D}} \label{Dyson}
\end{align}
where $\Sigma$ and $\gamma$ are the self-energy and three-leg
vertex, correspondingly, $\mathbf{K},\mathbf{P}$ are, as before,
$D$-dimensional wave vectors, and static correlation function is
given by the expression (\ref{statcorr}). To find asymptotic of
DOS for large enough negative energies one can use a method
proposed first by Keldysh for doped semiconductors \cite{efros}
(the same trick was used also for magnetic semiconductors near
$T_{c}$ \cite{bebe} and for electron topological transitions
\cite{ETT}). For large enough $\left| E\right|, E<0$ one can
neglect momentum dependence of both $\Sigma$ and $\gamma$ since
only the momentum transfer $\mathbf{K}\rightarrow0$ is relevant
for $d<2$. Also, we can express $\gamma$ in terms of $\Sigma$ via
the Ward identity \cite{migdal}
\begin{equation}
\gamma\left(\mathbf{P},\mathbf{P};0;E\right)
=1-\frac{\partial\Sigma\left( E\right)}{
\partial E}.  \label{ward}
\end{equation}
Then, taking into account Eq.(\ref{statcorr}), we obtain a closed
differential equation for the self-energy of the form
\begin{equation}
\Sigma\left( E\right) =\left( 1-\frac{\partial\Sigma\left( E\right) }{%
\partial E}\right) g^{2}A_{D}\int
_{0}^{\infty}\frac{K^{d-1}dK}{E-K^{2}/2-\Sigma\left( E\right) }.
\label{difeq}
\end{equation}

Consider now the density of states (DOS)
\begin{equation}
N_{D}\left( E\right) =-\frac{A_{D}}{\pi}\mbox{Im}
\int_{0}^{K_{\max}}\frac{K^{D-1}dK}{E -\Sigma\left(
K,E+i\delta\right) -\frac{1}{2}K^{2}}
\end{equation}
It is clear that at $\left| E-\Sigma\left( k,E+i\delta\right)
\right| \ll$ $\frac{1}{2}K_{\max}^{2}$ at least for $D\leq3$ the
main contribution to $N_{D}\left( E\right)$ comes from small $K$
($K\ll K_{\max } $) region.

Let us solve now the equation (\ref{difeq}). Integrating over $K$
one derives
\begin{equation}
\Sigma\left( E\right) =\frac{\pi D}{2\sin\frac{\pi
d}{2}}\frac{g^{2}}{W^{\frac{d}{2}}}\left[ \frac{d\Sigma\left(
E\right) }{dE}-1\right] \left[ \Sigma\left( E\right) -E\right]
^{\frac{d}{2}-1} \label{difeq1}
\end{equation}

Denoting
\begin{equation}
\Sigma\left( E\right) -E=E_{0}\left[ f\left(
\frac{E}{E_{0}}\right) \right] ^{\frac{2}{d}}  \label{E0sigma}
\end{equation}
with $E_0$ given by Eq.(\ref{E00}) we obtain a non-linear
first-order ordinary differential equation
\begin{equation}
\frac{2}{d}\frac{df}{dx}=f^{\frac{2}{d}}+x.  \label{difeq2}
\end{equation}
For $d=1$ this is Riccatti equation, which was solved in a similar
context earlier \cite{bebe}. We consider here only the asymptotic
behavior of the solution at $E<0$ and $\left| E\right| \gg E_{0}$
directly from the initial equation (\ref{difeq}). For these $E$
\begin{equation}
\left| \mbox{Im}\Sigma\left( E\right) \right| \ll\left| \mbox{Re}%
\Sigma\left( E\right) \right| \ll\left| E\right|  \label{siq113}
\end{equation}
and we linearize this equation with respect to the imaginary part
of the self-energy to obtain
\begin{equation}
\frac{d\mbox{Im}\Sigma\left( E\right)
}{dE}\simeq\frac{1}{E_{0}}\left( \frac{-E}{E_{0}}\right)
^{1-\frac{d}{2}}\mbox{Im}\Sigma\left( E\right) \label{sig114}
\end{equation}
Thus, we have
\begin{equation}
\mbox{Im}\Sigma\left( E\right) \simeq CE_{0}\exp\left[
-\frac{2}{4-d}\left| \frac{E}{E_{0}}\right|
^{2-\frac{d}{2}}\right] ,\;\left| E\right| \gg E_{0}
\label{sig115}
\end{equation}
where $C$ is an undetermined integration constant. At these
energies, the density of states becomes
\begin{equation}
N_{D}\left( E\right)  \simeq \frac{CD(2-D)}{2\sin\frac{\pi
D}{2}}\frac{E_{0}}{W^{\frac{D}{2}}\left| E\right|
^{2-\frac{D}{2}}}\exp\left[ -\frac{2}{4-d}\left|
\frac{E}{E_{0}}\right| ^{2-\frac{d}{2}}\right]
\end{equation}
which coincides with the result (\ref{density}), with an accuracy
of a numerical factor of order of 1 in the exponent. This may be
considered as a justification of our treatment basing on the
Feynman variational approach.

The physical meaning of the self-trapping energy for quantum and
classical fluctuons are essentially different. For the fluctuon
near QCP, as well as for the Feynman polaron, we calculate
approximately the ground state electron energy, or the edge of the
spectrum. If we will calculate next-order corrections to the
electron free energy in $T=\beta ^{-1}$ we will find just a
temperature shift of this energy rather than any exponential tail
of DOS. The energy of the classical fluctuon is just a position of
the chemical potential at small enough electron concentration $n$.
For
\begin{equation}
n<\int_{-\infty }^{0}N\left( E\right) dE\propto \left( \frac{g}{W}%
\right) ^{\frac{2D}{4-d}},  \label{tailcapac}
\end{equation}
which is a capacity of the tail, the chemical potential level is
``pinned'' to the fluctuon energy and almost independent on $n$
due to exponential dependence of the DOS (\ref{density}) on $E$.

\section{Conclusions}

Let us resume on the main results obtained. Due to complexity of
the problem of the electron states near quantum critical point
(QCP) it is hardly believable that this problem can be treated
rigorously. To obtain first insight into this we used variational
approach within Feynman path integral formalism. Originally, this
approach was developed in the connection with polaron in ionic
crystals and proved to give excellent results
\cite{feynman,path1}. For the case of classical critical point
(CCP) we have checked the reliability of this approach by fairly
independent Green function method.

The results on the electron ground state at QCP turn out to be
crucially dependent on the anomalous space dimensionality
$d=D-2+\eta$ and dynamical critical exponent $z$. The most
interesting result is nonexistence of regular perturbation theory
for the ground state energy for arbitrary small coupling constant
$g$. In such cases singular perturbation theory emerges with the
expansion in non-integer powers of $g$. For $z \geq 2$, those
cases fall into range $d+z-2 \leq 1$. For $z<2$ it occurs at $z
\geq \frac{2}{3}(d+2)$ which is consistent if $0<d<1$.

In the above mentioned singular perturbation-theory cases, as well
as in general situation at large enough $g$ (strong coupling
regime) the leading term in the ground state energy is independent
of $z$ and is given by Eq.(\ref{q_0, E_0sc}). This result is valid
for $g^2 \ll W\omega$ ($W$ is the electron bandwidth and $\omega$
is a typical fluctuation energy) which in fact is a criterion of
consistence of continuum approximation. Physically this means that
the size of self-trapped state (fluctuon) is much larger than
interatomic distance. Otherwise a small-radius fluctuon likely
forms, which should be considered by different methods.

In contrast with the quantum case, at CCP the fluctuon states form
a continuum in the DOS tail. In this case the variational
fluctuon's free energy by Feynman method simply gives a position
of the electron chemical potential in the tail counted from the
bare band edge. The tail capacity proves $\left( \frac{g}{W}
\right) ^{\frac{2D}{4-d}}$ times a numerical constant; if the
electron concentration is much larger than this estimate the
fluctuons can scarcely contribute to the electron properties of
material near CCP.

\end{document}